\documentclass[aps,amsmath,amssymb,twocolumn,showpacs]{revtex4-1}

\usepackage{graphicx}
\usepackage{dcolumn}
\usepackage{bm}
\usepackage{color}

\begin{document}

\title{Thermal conductance of metallic atomic-size contacts: \\ Phonon
  transport and Wiedemann-Franz law}

\author{J. C. Kl\"ockner$^{1}$}
\email{Corresponding author: Jan.Kloeckner@uni-konstanz.de}
\author{M. Matt$^{1}$}
\author{P. Nielaba$^{1}$}
\author{F. Pauly$^{1,2}$}
\author{J. C. Cuevas$^{1,3}$}

\affiliation{$^{1}$Department of Physics, University of Konstanz, D-78457 Konstanz, Germany}

\affiliation{$^{2}$Okinawa Institute of Science and Technology Graduate University, Onna-son, 
Okinawa 904-0395, Japan}

\affiliation{$^{3}$Departamento de F\'{\i}sica Te\'orica de la Materia Condensada and Condensed Matter
Physics Center (IFIMAC), Universidad Aut\'onoma de Madrid, E-28049 Madrid, Spain}

\date{\today}

\begin{abstract}
Motivated by recent experiments [Science {\bf 355}, 6330 (2017);
  Nat. Nanotechnol. {\bf 12}, 430 (2017)], we present here an extensive
theoretical analysis of the thermal conductance of atomic-size contacts made
of three different metals, namely gold (Au), platinum (Pt) and aluminum
(Al). The main goal of this work is to elucidate the role of phonons in the
thermal transport through these atomic contacts as well as to
study the validity of the Wiedemann-Franz law, which relates the electrical
and the thermal conductance. For this purpose, we have employed two different
custom-developed theoretical approaches. The first one is a transport method
based on density functional theory (DFT) that allows to accurately compute the
contributions of both electrons and phonons to the thermal transport in
few-atom-thick contacts. The second technique is based on a combination of
classical molecular dynamics (MD) simulations and a tight-binding model that
enables the efficient calculation of the electronic contribution to the
thermal conductance of atomic contacts of larger size. Our DFT-based
calculations show that the thermal conductance of few-atom contacts of Au and
Pt is dominated by electrons, with phonons giving a contribution typically
below 10\% of the total thermal conductance, depending on the contact
geometry.  For these two metals we find that the small deviations from the
Wiedemann-Franz law, reported experimentally, largely stem from phonons. In 
the case of Al contacts we predict that the phononic contribution can be 
considerably larger with up to 40\% of the total thermal conductance. We 
show that these differences in the phononic contribution across metals 
originate mainly from their distinct Debye energies. On the other hand, 
our MD-based calculations demonstrate that the electronic contribution to 
the thermal conductance follows very closely the Wiedemann-Franz law, 
irrespective of the material and the contact size. Finally, the ensemble 
of our results consistently shows that the reported observation of 
quantized thermal transport at room temperature is restricted to few-atom 
contacts of Au, a monovalent metal in which the transport is dominated 
by the $s$ valence orbitals. In the case of multivalent metals like Pt 
and Al this quantization is statistically absent due to the fact that 
additional orbitals contribute to the transport with conduction channels 
that have intermediate transmissions between 0 and 1, even in the case of 
single-atom contacts.
\end{abstract}

\maketitle

\section{Introduction}
The advent of experimental techniques like the scanning tunneling microscope
and the mechanically controllable break-junctions made it possible in the
early 1990s to fabricate metallic atomic-size contacts all the way down to
single-atom junctions and even chains of atoms \cite{Agrait2003,Cuevas2017}.
This paved the way for investigating a large variety of charge and energy
transport properties in these atomic contacts such as electrical conductance
\cite{Krans1995,Scheer1998}, shot noise
\cite{Brom1999,Wheeler2010,Chen2012,Vardimon2016}, photocurrent
\cite{Guhr2007,Viljas2007,Ittah2009,Ward2010}, thermopower
\cite{Ludoph1999,Tsutsui2013,Evangeli2015,Ofarim2016}, and Joule heating
\cite{Lee2013,Zotti2014}, just to mention a few. The mean free
  path for electrons and phonons is larger than the characteristic dimensions
  of atomic contacts \cite{Ashcroft1976}, even at room temperature, and all
the transport properties of these nanowires are therefore dominated by quantum
mechanical effects. For this reason, metallic atomic-size contacts have become
an ideal playground to test basic quantum theories of charge and energy
transport at the nanoscale. In fact, one can safely say that no other system
has contributed so decisively to firmly establish the quantum coherent
transport picture put forward by Rolf Landauer, Markus B\"uttiker, Joseph Imry
and others \cite{Agrait2003,Cuevas2017,Buttiker1985,Buttiker1986,Imry1999}.

Until recently, there was a basic transport property that had not been
experimentally investigated in metallic atomic-size contacts, namely the
thermal conductance. This situation has now changed, and two experimental
groups have finally been able to independently explore the thermal transport
in atomic contacts \cite{Cui2017,Mosso2017}. In particular, Cui \emph{et
  al.}\ \cite{Cui2017} were able to measure the room-temperature thermal
conductance of both Au and Pt contacts and found that in the case of Au
single-atom contacts the thermal conductance is quantized in units of the
universal thermal conductance quantum $\kappa_0 = \pi^2 k^2_{\rm B} T/(3h)$,
where $T$ is the absolute temperature. Before this, thermal conductance
quantization had already been reported in a series of experiments, making use
of different microdevices, where the heat was carried by phonons, electrons or
photons
\cite{Schwab2000,Chiatti2006,Meschke2006,Jezouin2013,Partanen2016}. But in all
cases sub-Kelvin temperatures were a necessary prerequisite for the
observation of this quantum phenomenon. Thus, the results of Cui \emph{et
  al.}\ \cite{Cui2017} constitute the first observation of quantized thermal
transport at room temperature and demonstrate the potential of these atomic
contacts to reveal novel quantum effects in thermal transport. The experiments
of Ref.~\onlinecite{Cui2017} also show that the thermal conductance of Pt
atomic-size contacts is not quantized, which is indeed expected in view of the
fundamentally different electronic structure of this metal as compared to Au.

Another important observation of Refs.~\onlinecite{Cui2017,Mosso2017} was that
the thermal conductance of atomic contacts follows closely the Wiedemann-Franz
law, irrespective of the material and the contact size. This law establishes
that the thermal conductance of a metal, where the heat transport is dominated
by electrons, is simply proportional to its electrical conductance. The
relation is known to be approximately fulfilled in macroscopic metallic wires
made of standard metals, and in this case it can be explained with the help of
the semiclassical Boltzmann transport equation \cite{Ziman2001}. For nanoscale
devices, where the transport is fully coherent, the validity of the
Wiedemann-Franz law requires two things: (i) The electronic transmission
function must be rather smooth around the Fermi energy in an energy window of
the thermal broadening $k_{\text{B}}T$
\cite{Butcher1990,vanHouten1992,Cuevas2017}, and (ii) the thermal transport
must be largely dominated by electrons \cite{Note1}.  The first condition is
indeed expected to be met by metallic atomic-size contacts, irrespective of
the material and the contact size, as shown by numerous investigations of the
conductance \cite{Agrait2003} and thermopower \cite{Ludoph1999,Evangeli2015}
of these nanowires. The second condition, which is also necessary for the
observation of quantized thermal transport in Ref.~\onlinecite{Cui2017}, is by
no means trivial. It is well known that the thermal conductance of macroscopic
metallic wires is largely dominated by electrons, with the phonons giving a
contribution that amounts to only a few percent of the total thermal
conductivity \cite{Jain2016}. However, when going from the macro- to the
nanoscale, the phonon transport mechanism changes from incoherent to coherent,
and it is not obvious a priori, whether the phonon contribution to the thermal
transport of metallic atomic-size contacts is actually negligible. In fact,
Refs.~\onlinecite{Cui2017,Mosso2017} reported slight
deviations from the Wiedemann-Franz law that were actually attributed to the
contribution of phonons and possibly to a small one of photons (thermal
radiation) \cite{Cui2017}. In this respect, we already showed in
Ref.~\onlinecite{Cui2017} that the contribution of the phonons to the thermal
conductance of Au single-atom contacts is only around 5\% of the total one,
which explains the validity of the Wiedemann-Franz law and, in turn, the
observation of quantized transport in these contacts.  The main goal of this
work is to provide a comprehensive analysis of the phonon transport in
metallic atomic-size contacts made not only of Au, but also of other relevant
metals like Pt and Al.

In this work we aim at elucidating the magnitude of the phonon contribution to
the thermal conductance of metallic atomic-size contacts and to shed light on
to what extent the Wiedemann-Franz law is expected to be fulfilled. For this
purpose we have made use of different custom-designed theoretical techniques
to describe the thermal transport in atomic junctions. In particular we have
employed a full ab initio, DFT-based transport method to compute the
contributions of both electrons and phonons to the thermal conductance of
atomic contacts made of Au, Pt, and Al. Our calculations show that, depending
on the contact size and the exact geometry, the phonons contribute by about
5-10\% to the total room-temperature thermal conductance of Au and Pt
contacts, which explains the small deviations from the Wiedemann-Franz law
observed in Ref.~\onlinecite{Cui2017}. Aluminum is a light metal with a high Debye
energy of about 40 meV as opposed to the 20 and 25 meV of Au and Pt,
respectively. In this case we calculate that the phonons can constitute up to
40\% of the total thermal conductance contribution, depending on the
geometry. These results show that phonons can in general not be ignored in the
analysis of the thermal transport through metallic atomic-size contacts,
especially for light metals. This is at variance with the case of macroscopic
metallic wires \cite{Jain2016}.

In addition we have made use of a combination of classical MD simulations
together with quantum mechanical calculations of the electronic thermal
conductance based on a tight-binding model to carry out a detailed study of
the validity of the Wiedemann-Franz law for Au, Pt, and Al contacts. Although
neglecting the phonon contribution to the thermal conductance, the strength of
this method is that it allows us to perform a detailed statistical analysis
and to simulate both the electrical and electronic thermal conductance
histograms, which enables us to establish a very direct comparison with the
experiments. Our analysis confirms the expectation that the electronic thermal
conductance fulfills in a very accurate manner the Wiedemann-Franz law for all
three metals and irrespective of the contact size. As compared
to Au and Al, we find slightly increased deviations of the order of up to
5\% in the case of Pt contacts due to the fact that the transport in this
metal is dominated by $d$ orbitals.

The rest of the paper is organized as follows. In the next section
\ref{sec-Methods} we describe the different theoretical techniques that we
have used to simulate the thermal transport of metallic atomic-size contacts.
In particular, we present in section \ref{sec-LB} the basic formulas that
describe the quantum thermal transport in these systems within the
Landauer-B\"uttiker approach. Then, in section \ref{sec-DFT-results} we
discuss the main results obtained with our DFT-based transport method for the
thermal conductance of Au, Pt, and Al few-atom contacts, with special emphasis
on the relative contributions of electrons and phonons. Section
\ref{sec-TB-results} is devoted to the analysis of the results for the
electronic contribution to the thermal conductance of Au, Pt, and Al contacts,
obtained with the help of the combined MD and tight-binding
simulations. Finally, in section \ref{sec-Conclusions} we summarize the main
conclusions of this work.

\section{Theoretical approaches} \label{sec-Methods}

In this section we describe the thermal conductance of metallic atomic-size
contacts within the Landauer-B\"uttiker formalism for coherent quantum
transport. For this purpose, we combine a number of theoretical techniques,
namely the non-equilibrium Green's function (NEGF) formalism, various
electronic structure methods and classical MD simulations. We will describe
these different theoretical methods in certain detail. To be precise, we will
first summarize the basic formulas of the Landauer-B\"uttiker approach to
compute the thermal conductance of a nanoscale system. Then, we will introduce
our DFT-based transport approach that allows us to compute both the electronic
and the phononic contributions to the thermal conductance. Finally, we will
describe the MD simulations that we use to determine the geometries of the
atomic contacts and how we combine the MD with a tight-binding model to
compute the electronic contribution to the different transport properties of
these atomic-scale contacts.

\subsection{Thermal conductance within the Landauer-B\"uttiker approach} \label{sec-LB}

There are two basic contributions to the thermal conductance of an atomic-size
contact, namely those of electrons and phonons. The inelastic scattering
lengths for electrons and phonons are clearly larger than the typical size of
these metallic atomic contacts \cite{Ashcroft1976}, even at room
temperature. Thus the description of the transport properties of these systems
can be performed within the Landauer-B\"uttiker approach for coherent
transport, where one assumes that the transport, both electronic and phononic,
is dominated by elastic scattering. Within this approach the contributions of
electrons and phonons to the different transport properties are determined by
the corresponding electronic and phononic transmission functions $\tau_{\rm
  el}$ and $\tau_{\rm ph}$, respectively. The electronic contribution to the
linear thermal conductance $\kappa_{\rm el}$ is given by
\cite{Cuevas2017,Sivan1986}
\begin{equation}
\kappa_{\rm el} = \frac{2}{hT} \left( K_{2}- \frac{K_{1}^{2}}{K_{0}}\right), 
\label{eq-kel}
\end{equation}
where $T$ is the absolute temperature and the $K_n$ coefficients are defined as
\begin{equation}
K_n = \int^{\infty}_{-\infty} (E-\mu)^{n} \tau_{\rm el}(E) \left(-\frac{\partial f(E,T)}
{\partial E}\right) dE.
  \label{eq-Kn}
\end{equation}
Here $f(E,\mu,T) = \left\{ \exp[(E-\mu)/k_{\mathrm{B}}T]+1\right\} ^{-1}$ is
the Fermi function, and the chemical potential $\mu\approx E_{\textnormal{F}}$
is approximately given by the Fermi energy $E_{\rm F}$ of the electrodes. At
this point it is important to notice that, if the electronic transmission does
not strongly depend on energy in the range of some $k_{\text{B}}T$ around
$E_{\rm F}$, the electronic thermal conductance is approximately given by the
Wiedemann-Franz law, i.e.,
\begin{equation}
\kappa_{\rm el} \approx L_0 T G.
\label{eq-WF-law}
\end{equation}
Here $L_0 = \pi^2 k^2_{\rm B}/(3e^2) = 2.44 \times 10^{-8}$ W$\Omega$K$^{-2}$
is the so-called Lorentz number and $G$ is the electrical conductance, which
we calculate as \cite{Cuevas2017}
\begin{equation}
  G = G_0 K_0,
\end{equation}
 with $K_0$ defined via Eq.~(\ref{eq-Kn}). For low temperatures, $G$ reduces
 to $G = G_0 \tau_{\rm el}(E_{\rm F})$, where $G_0 = 2e^2/h = 77.48$ $\mu$S is
 the electrical conductance quantum. The test of the validity of this law in
 metallic atomic-size contacts is one of the central issues of this
 work. Notice that Eq.~(\ref{eq-WF-law}) can be rewritten as $\kappa_{\rm el}
 = 2 \kappa_0 \tau_{\rm el}(E_{\rm F})$, where $\kappa_0 = \pi^2 k^2_{\rm B}
 T/(3h)$ is the thermal conductance quantum, which takes the value of
 $\kappa_0 \approx 0.284$~nW/K at room temperature ($T=300$~K). The factor 2 in
 the previous equation is related to the spin degeneracy that is assumed in
 this work for the electronic transport. Thus, we see that in a system, where
 electrons dominate the quantum transport and the Wiedemann-Franz law is
 fulfilled, the quantization of the electrical conductance implies the
 quantization of the thermal conductance.
  
The corresponding phonon thermal conductance in the linear response regime is
given by \cite{Rego1998,Mingo2003,Yamamoto2006}
\begin{equation}
\kappa_{\rm ph} = \frac{1}{h} \int_{0}^{\infty} E \tau_{\rm ph}(E) 
\frac{\partial n(E,T)}{\partial T} dE, 
\label{eq-kph}
\end{equation}
where $n(E,T)=[\exp(E/k_{\rm B}T)-1]^{-1}$ is the Bose function, describing
the phonon occupation in the electrodes. To get an idea about the order of
magnitude of the phononic thermal conductance, we can express the previous
equation as follows
\begin{equation}
\kappa_{\rm ph} = \kappa_0 \int_{0}^{\infty} W_{\rm ph}(E,T) \tau_{\rm ph}(E) dE, 
\label{eq-kph2}
\end{equation}
where $\kappa_0$ is the thermal conductance quantum introduced above and the
``window'' function $W_{\rm ph}(E,T)$ is defined as
\begin{equation}
W_{\rm ph}(E,T) = \frac{3}{\pi^2} \left( \frac{E}{k_{\rm B} T}\right)^2 
\left( - \frac{\partial n(E,T)}{\partial E} \right).  
\label{eq-Wph}
\end{equation}
It fulfills the normalization condition
\begin{equation}
  \int_{0}^{\infty} W_{\rm ph}(E,T) dE = 1.
\end{equation}
Thus, if we assume that $\tau_{\rm ph}(E) = 1$ over the whole energy range,
over which the function $W_{\rm ph}$ has a sizable value, then $\kappa_{\rm ph} 
= \kappa_0$. As we shall see below (c.f.\ Fig.~\ref{fig-estimate}), this 
condition is difficult to fulfill at room temperature due to the finite Debye 
energy of the different metals.

The bottom line of the previous discussion is that the description of the
different transport properties, investigated in this work, requires the
calculation of the electronic and phononic transmission functions. In the
following subsections we will show, how we compute these functions with the
help of the NEGF technique and different electronic structure methods.

\subsection{DFT-based transport calculations}

In this subsection we describe, how we combine density functional theory (DFT)
with NEGF techniques to compute the electrical and thermal conductance of
atomic-size contacts, taking into account the contributions of electrons and 
phonons. This combination makes use of the first-principles formalism developed 
by some of us and reported in Refs.~\onlinecite{Pauly2008,Burkle2015}. In what 
follows, we briefly describe this formalism.
   
\subsubsection{Contact geometries, electronic structure, and vibrational
  properties} \label{sec-CGESVP}

The first step in our ab initio calculations is the construction of the atomic
junction geometries. As described below in more detail, we investigate ideal
geometries to simulate one-atom-thick contacts and study also the stretching
of the atomic junctions to determine conductance traces in the spirit of the
experiments of Ref.~\onlinecite{Cui2017}. In both cases we make use of DFT to
compute equilibrium geometries through total energy minimization and to
describe their electronic structure. Vibrational properties of the optimized
contacts are obtained in the framework of density functional perturbation
theory (DFPT).

We use both DFT and DFPT procedures, as implemented in the quantum chemistry 
software package TURBOMOLE 6.5 \cite{TURBOMOLE,Deglmann2002,Deglmann2004}. In 
our calculations we employ the PBE exchange-correlation functional
\cite{Perdew1992,Perdew1996}, the basis sets are def2-SV(P)
for Au \cite{Weigend2005} and def-TZVP for Pt and Al \cite{Schafer1994}. Due 
to linear dependencies in bulk calculations of Pt, we have changed the exponent 
of the most diffuse $s$ basis function from $0.04$~a.u.$^{-2}$ to $0.07$~a.u.$^{-2}$ 
and those of the most diffuse $p$ function from $0.05$~a.u.$^{-2}$ to $0.08$~a.u.$^{-2}$. 
In all cases the corresponding Coulomb fitting basis is employed 
\cite{Eichkorn1997,Weigend2006}. To ensure that the vibrational properties, i.e., 
force constants and derived vibrational energies, are accurately determined, we 
use very stringent convergence criteria to avoid the appearance of imaginary
frequencies in the optimized contact region. Thus, total energies are converged 
to a precision of better than $10^{-9}$~a.u., while geometry optimizations are 
continued until the change of the maximum norm of the Cartesian gradient is 
below $10^{-5}$~a.u.

The bulk phonon properties of Au are determined as described in
Ref.~\onlinecite{Burkle2015}.  In short the dynamical matrix of the Au
electrode is derived from those of a spherical cluster. As in the crystal the
atoms of the Au cluster are positioned on an fcc lattice with a lattice
constant of $a_{\text{Au}}=4.08$~\AA, and force constants from the central
atom to its neighbors are extracted from a sufficiently large cluster of 333
atoms. Since phonon properties are easily exchangeable between different
electronic structure codes and in order to avoid the calculation of large
clusters, we have changed our computational strategy for Pt and Al. For these
two materials we use DFT and DFPT, as implemented in the plane wave code
QUANTUM ESPRESSO \cite{Giannozzi2009} with PAW pseudopotentials taken from the
PS Library \cite{DalCorso2012}, to calculate bulk force
  constants and phonon properties. For Pt we employ a grid of $24 \times 24
\times 24$ electronic $k$-points, an energy cutoff of 100 Ry and a
Marzari-Vanderbilt smearing of 0.07 Ry. The phonons are then computed on a
grid of $9 \times 9 \times 9$ $q$-points. Similarly, for Al we utilize $24
\times 24 \times 24$ $k$-points, an energy cutoff of 100 Ry, a
Marzari-Vanderbilt smearing of 0.1 Ry and a $q$-grid of $11 \times 11 \times
11$. Similar to Au we consistently use the experimental lattice constants
$a_{\rm Pt} = 3.92$ \AA\ and $a_{\rm Al} = 4.05$ \AA\ for the calculation of
the bulk properties \cite{Ashcroft1976}.

\subsubsection{Electron transport}

To determine the electronic structure of the atomic junctions
and to compute the electronic transmission that fixes the electrical
conductance and the electronic contribution to the thermal conductance
within the Landauer-B\"uttiker approach, we use NEGFs expressed in a local
non-orthogonal basis. Briefly, the local basis allows us to partition the
basis states into L, C and R ones, where L and R correspond to the left and
right electrode, respectively, while C corresponds to a central region
including the atomic neck. Thus, the (single-particle) Hamiltonian (or Fock)
matrix $\boldsymbol H$ can be written in the block form
\begin{equation}
{\boldsymbol H} = \left(\begin{array}{ccc}
{\boldsymbol H}_{\rm LL} & {\boldsymbol H}_{\rm LC} & {\boldsymbol 0}\\ {\boldsymbol H}_{\rm CL} & 
{\boldsymbol H}_{\rm CC} & {\boldsymbol H}_{\rm CR}\\ {\boldsymbol 0} & {\boldsymbol H}_{\rm RC} & 
{\boldsymbol H}_{\rm RR} \end{array}\right).
\label{eq-H}
\end{equation}
A similar expression holds for the overlap matrix $\boldsymbol S$. The
energy-dependent electronic transmission $\tau_{\rm el}(E)$ can be expressed
in terms of the Green's functions as \cite{Cuevas2017}
\begin{equation}
\tau_{\rm el}(E) = \mathrm{Tr} \left[{\boldsymbol \Gamma}_{\rm L}(E) {\boldsymbol G}_{\rm CC}^{\rm r}(E) 
{\boldsymbol \Gamma}_{\rm R}(E) {\boldsymbol G}_{\rm CC}^{\rm a}(E) \right],
\label{eq-tau-el}
\end{equation}
where the retarded Green's function is given by
\begin{equation}
{\boldsymbol G}_{\rm CC}^{\rm r}(E) = \left[ (E+i\eta) {\boldsymbol S}_{CC} - {\boldsymbol H}_{CC} - 
{\boldsymbol \Sigma}_{L}^{r}(E) - {\boldsymbol \Sigma}_{R}^{r}(E)\right]^{-1} .
\label{eq-GCC}
\end{equation}
Here, $\eta$ is an infinitesimal positive parameter (that will be omitted
hereafter), and advanced and retarded Green's functions are related by
${\boldsymbol G}_{\rm CC}^{\rm a} = \left[{\boldsymbol G}_{\rm CC}^{\rm
    r}\right]^{\dagger}$. The retarded self-energies in the previous equation
adopt the form
\begin{equation}
{\boldsymbol \Sigma}_{X}^{\rm r}(E) = \left({\boldsymbol H}_{{\rm C}X} - E {\boldsymbol S}_{{\rm C}X} 
\right) {\boldsymbol g}_{XX}^{\rm r}(E) \left({\boldsymbol H}_{X{\rm C}} - E {\boldsymbol S}_{X{\rm C}} 
\right).
\label{eq-SigmaX}
\end{equation}
The scattering rate matrices that enter the expression of the electronic
transmission are given by ${\boldsymbol \Gamma}_{X}(E) = i \left[ \boldsymbol
  \Sigma_{X}^{\rm r}(E) - \boldsymbol \Sigma_{X}^{\rm a}(E) \right]$, and
${\boldsymbol g}_{XX}^{\rm r}(E) = (E {\boldsymbol S}_{XX} - {\boldsymbol
  H}_{XX})^{-1}$ are the electrode Green's functions with $X=\mbox{L,
  R}$. Finally, it is convenient to decompose the total electronic
transmission in terms of individual transmission coefficients. For this
purpose we can write Eq.~(\ref{eq-tau-el}) as
\begin{equation}
\tau_{\rm el}(E) = \mathrm{Tr} \left[ \boldsymbol t_{\rm el}(E) \boldsymbol 
t^{\dagger}_{\rm el} (E) \right] = \sum_i \tau_{{\rm el},i}(E) ,
\label{eq-tau-chan}
\end{equation}
where $\boldsymbol t_{\rm el}(E) = \boldsymbol \Gamma^{1/2}_{\rm L}(E)
\boldsymbol G_{\rm CC}^{\rm r}(E) \boldsymbol \Gamma^{1/2}_{\rm R}(E)$ is the
electronic transmission amplitude matrix and $\tau_{{\rm el},i}(E)$ are the
eigenvalues of the transmission probability matrix $\boldsymbol t_{\rm el}(E)
\boldsymbol t^{\dagger}_{\rm el}(E)$. They are known as transmission
coefficients, while the corresponding eigenfunctions are referred to as
conduction channels.

In order to describe the transport through the atomic contacts, we first
extract $\boldsymbol H_{\rm CC}$ and $\boldsymbol S_{\rm CC}$ and the matrices
$\boldsymbol H_{{\rm C}X}$ and $\boldsymbol S_{{\rm C}X}$ from a DFT
calculation of an extended central cluster that includes the central wire and
part of the leads.  On the other hand, the electrode Green's functions
$\boldsymbol g_{XX}^{\rm r}(E)$ are modeled as surface Green's functions of
ideal semi-infinite crystals. To obtain these Green's functions, we first
compute separately the electronic structure of a large spherical fcc cluster 
of 1415 atoms. Then we extract the bulk Hamiltonian and overlap matrix elements, 
and we use them to model a semi-infinite crystal that is infinitely extended
perpendicular to the transport direction. The surface Green's functions are
calculated from this crystal with the help of a decimation technique
\cite{Pauly2008,Guinea1983}. In this way we describe the whole system
consistently within DFT, using the same nonorthogonal basis set and
exchange-correlation functional everywhere.

\subsubsection{Phonon transport}

To compute the phonon transmission, appearing in Eq.~(\ref{eq-kph}), we use 
our previous work \cite{Burkle2015,Klockner2016,Cui2017,Klockner2017a} and
combine DFT and NEGF techniques in the same spirit as for the
electron transport. Briefly our starting point is the description of the
phonons or vibrational modes of the atomic contacts within the harmonic
approximation. In this approximation the phonon Hamiltonian for small
displacements $\{Q_{\xi}\}$ of the atoms around their equilibrium positions
$\{R_{\xi}^{(0)}\}$ adopts the form
\begin{equation}
  \hat H = \frac{1}{2} \sum_{\xi} \hat{p}_{\xi}^{2} + \frac{1}{2\hbar^{2}} 
  \sum_{\xi\chi} \hat{q}_{\xi} K_{\xi\chi} \hat{q}_{\chi},
\end{equation}
where we have introduced mass-weighted displacement operators $\hat{q}_{\xi} =
\sqrt{M_{\xi}}\hat{Q}_{\xi}$ and mass-scaled momentum operators
$\hat{p}_{\xi}=\hat{P}_{\xi}/\sqrt{M_{\xi}}$ as conjugate variables.  These
variables obey the following commutation relations:
$[\hat{q}_{\xi},\hat{p}_{\chi}] = \mbox{i}\hbar\delta_{\xi\chi}$ and
$[\hat{q}_{\xi},\hat{q}_{\chi}] = [\hat{p}_{\xi},\hat{p}_{\chi}]=0$. Here
$\xi=(j,c)$ denotes a Cartesian component $c=x,y,z$ of atom $j$ at position
$\vec{R}_{j} = \vec{R}_{j}^{(0)}+\vec{Q}_{j}$. The phonon system is
characterized by its dynamical matrix $K_{\xi\chi} = \hbar^{2}
\partial_{\xi\chi}^{2}E_{\rm DFT}/\sqrt{M_{\xi}M_{\chi}}$, which is the
mass-weighted Hessian of the DFT total ground state energy $E_{\rm DFT}$ with
respect to the Cartesian atomic coordinates. These harmonic force constants
are computed within DFPT.

In analogy with the electronic system above, the use of a local displacement
basis enables the partitioning of the dynamical matrix into three parts, a
central scattering region C, and the two semi-infinite L and R electrodes
\begin{equation}
  \boldsymbol{K}=\left(\begin{array}{ccc}
    \boldsymbol{K}_{\mathrm{LL}} & \boldsymbol{K}_{\mathrm{LC}} & \mathrm{\boldsymbol{0}}\\
    \boldsymbol{K}_{\mathrm{CL}} & \boldsymbol{K}_{\mathrm{CC}} & \boldsymbol{K}_{\mathrm{CR}}\\
    \boldsymbol{0} & \boldsymbol{K}_{\mathrm{RC}} & \boldsymbol{K}_{\mathrm{RR}}
  \end{array}\right). 
\label{eq-Kmatrix}
\end{equation}
The energy-dependent phononic transmission $\tau_{\rm ph}(E)$ can be expressed in terms of 
phonon Green's functions as \cite{Mingo2003,Burkle2015}
\begin{equation}
\tau_{\rm ph}(E) = \mathrm{Tr} \left[ \boldsymbol{D}_{\mathrm{CC}}^{\textrm{r}}(E) 
 \boldsymbol{\Lambda}_{\textrm{L}}(E) \boldsymbol{D}_{\mathrm{CC}}^{\textrm{a}}(E)
 \boldsymbol{\Lambda}_{\textrm{R}}(E)\right],
\label{eq-tauph}
\end{equation}
where $\boldsymbol{D}_{\mathrm{CC}}^{\textrm{r,a}}(E)$ are the retarded and
advanced phonon Green's functions of the central region. They can be computed
by solving the following Dyson equation
\begin{equation}
  \boldsymbol{D}_{\mathrm{CC}}^{\mathrm{r}}(E) = \left[\left(E+i\eta
    \right)^{2}\boldsymbol{1}_{\textrm{CC}} - \boldsymbol{K}_{\textrm{CC}} -
    \boldsymbol{\Pi}_{\textrm{L}}^{\mathrm{r}}(E) -
    \boldsymbol{\Pi}_{\textrm{R}}^{\mathrm{r}}(E) \right]^{-1} .
\end{equation}
In the expression, $\eta>0$ is an infinitesimal parameter and
$\boldsymbol{D}_{\rm CC}^{\rm a}(E) = \boldsymbol{D}_{\rm CC}^{\rm
  r}(E)^{\dagger}$. The scattering rate matrices
\begin{equation}
  \boldsymbol{\Lambda}_{X}(E)= i
  \left[\boldsymbol{\Pi}_{X}^{\mathrm{r}}(E)-\boldsymbol{\Pi}_{X}^{\mathrm{a}}(E)\right]
\end{equation}
are related to the corresponding embedding self-energies
\begin{equation}
  \boldsymbol{\Pi}_{X}^{\mathrm{r}}(E) = \boldsymbol{K}_{\textrm{C}X} 
  \boldsymbol{d}_{XX}^{\mathrm{r}}(E) \boldsymbol{K}_{X\mathrm{C}},
\end{equation}
which describe the coupling between the central region C and electrode $X$. In
the expressions $\boldsymbol{d}_{XX}^{\rm r}(E)= [ (E+i\eta)^{2}\boldsymbol{1}_{XX} - 
\boldsymbol{K}_{\textrm{XX}}]^{-1}$ is the surface Green's function of lead 
$X=\text{L},\text{R}$, and $\boldsymbol{\Pi}_{X}^{\textrm{a}}(E) = 
\boldsymbol{\Pi}_{X}^{\textrm{r}}(E)^{\dagger}$.

To calculate the different parts of the dynamical matrix in
Eq.~(\ref{eq-Kmatrix}) we follow the same strategy as in the electronic case
described above. We first compute the dynamical matrix for an extended central
cluster, including the atomic wire and parts of the leads.  Subsequently we
extract from it the matrices $\boldsymbol K_{\rm CC}$ and $\boldsymbol K_{{\rm
    C}X}$.  On the other hand, the surface Green's functions of the electrodes
$\boldsymbol d_{XX}^{\rm r}(E)$ are obtained by extracting bulk force
constants either from a separate calculation of a big cluster or from a
periodic bulk calculation, as discussed before in subsubsection~\ref{sec-CGESVP}. 
The extracted bulk parameters are then used in combination with a decimation technique
\cite{Pauly2008,Guinea1983} to describe the surface of a semi-infinite perfect
crystal, exactly like in the electronic case.

Let us close this subsubsection by pointing out another
analogy with the electronic case: We can decompose the total phononic
transmission of Eq.~(\ref{eq-tauph}) into the contribution of individual
phonon transmission coefficients, $\tau_{\rm ph}(E) = \sum_i \tau_{{\rm
ph},i}(E)$. Here, the coefficients $\tau_{{\rm ph},i}(E)$ are the
eigenvalues of $\boldsymbol t_{\rm ph}(E) \boldsymbol t^{\dagger}_{\rm
ph}(E)$, where $\boldsymbol t_{\rm ph}(E) = \boldsymbol \Lambda^{1/2}_{\rm
L}(E) \boldsymbol D_{\rm CC}^{\rm r}(E) \boldsymbol \Lambda^{1/2}_{\rm
R}(E)$ is the phononic transmission amplitude matrix.

\subsection{Combination of MD simulations and a tight-binding model} \label{sec-MD}

The DFT-based transport calculations are very time-consuming and to carry out
a complete statistical analysis within this ab initio approach is out of reach
at present. Instead we are able to perform such an analysis for the electronic
transport (both for $G$ and $\kappa_{\text{el}}$) by combining classical MD
simulations of the junction formation with quantum transport calculations
based on a sophisticated tight-binding model. In particular, this hybrid
approach allows us to compute conductance histograms that can be directly
compared with those reported experimentally. Indeed this combination has been
quite successful in determining a variety of properties of these atomic-scale
wires
\cite{Dreher2005,Pauly2006,Pauly2011,Schirm2013,Chen2014,Evangeli2015,Vardimon2016,Cui2017}. The
disadvantage of this method, as compared to our DFT-based approach, is that at
present we are not able to describe the phonon transport. But since the
phonons will be shown to play a minor role in most situations, calculations of
this type are extremely useful to explore, in particular, the validity of the
Wiedemann-Franz law in metallic atomic-size contacts of several hundred atoms
in size. In what follows, we shall describe this combined approach by
presenting separately the details of the MD simulations and those related to
the transport calculations based on the tight-binding model.

\subsubsection{Molecular dynamics simulations}

\begin{figure}[t]
\begin{center} \includegraphics[width=0.8\columnwidth,clip]{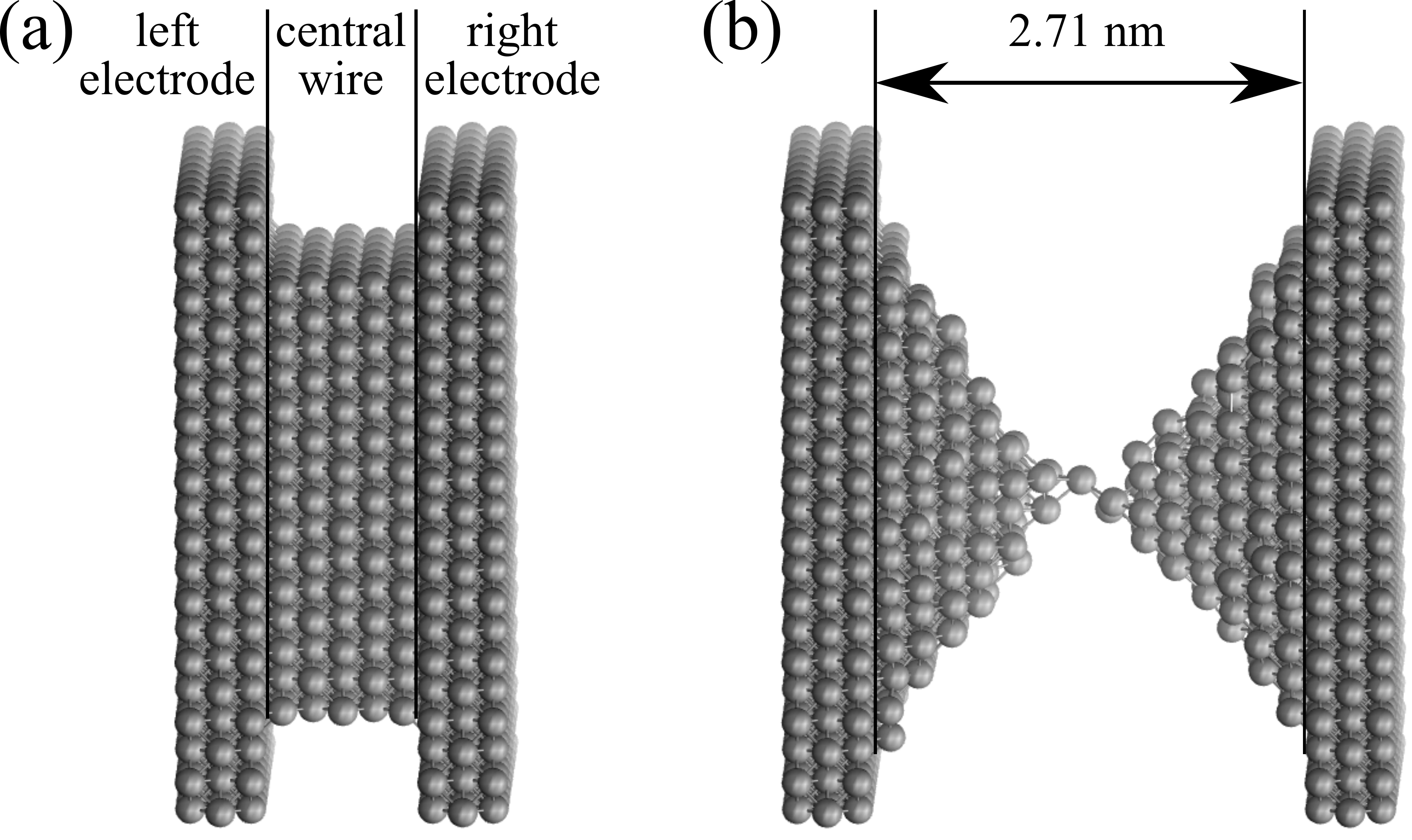} \end{center}
\caption{(Color online) (a) The initial fcc structure of the atomic contacts
  employed in the MD simulations. The example is for Al, where the central
  wire posses a length of 0.84 nm. We also show the partitioning of the
  contact into the left and right electrodes and the central wire, as used for
  the MD and transport calculations. (b) Example of atomic contacts of Al at
  an elongation of 1.87 nm. Considering the initial length of the junction of
  0.84 nm, the total length between of the central wire amounts to 2.71 nm, as
  indicated in the panel.}
\label{fig-geometry}
\end{figure}

As mentioned above, in order to perform a thorough statistical analysis of the
thermal conductance of different metallic atomic-size contacts, we carry out
classical MD simulations to first determine the geometry of these contacts
following our previous work \cite{Evangeli2015}. The thermal conductance is
then computed for the geometries determined with these MD simulations, as we
explain in the next paragraph.  We perform the MD simulations with the open
source program package LAMMPS \cite{LAMMPS,Plimpton1995}.  Within LAMMPS we
use the embedded atom method with the semi-empirical potentials from Ackland
\emph{et al.}\ \cite{Ackland1987} for Au and from Sheng \emph{et
  al.}\ \cite{Sheng2011} for Pt and Al to model the interactions between
atoms. It is worth stressing that these potentials account for the possibility
to have an atomic coordination that differs from the bulk. In order to obtain
the geometry of the atomic contacts, we start with an ideal fcc lattice, where
the crystal direction $\langle 100 \rangle$ lies parallel to the transport and
elongation direction. In these simulations we first divide the geometry into
three parts: Two electrodes and a central wire, bridging the gap between them,
as shown in Fig.~\ref{fig-geometry}. Each electrode consists of 661 atoms that
are kept fixed during the simulations. The size of the
  electrodes is chosen such that they contain at least all those atoms that
  are separated from the central wire atoms by less than the cut-off radius of
  the interaction potentials. The central wire is made up of 563 atoms and
their motion is described with Newtonian equations of motion. We assume a
canonical ensemble and use the velocity Verlet integration scheme
\cite{Frenkel2002}.  In our simulations the wires have an initial length of
0.82 nm for Au, 0.78 nm for Pt, and 0.84 nm for Al. The starting velocities of
the atoms in the wire are chosen randomly with a Gaussian distribution to
yield an average temperature of $T = 300$~K. Because of the randomness in the
initial velocity distribution, every stretching simulation is different, while
a Nos\'e-Hoover thermostat ensures that the temperature remains fixed
\cite{Frenkel2002}. The time step in all our simulations is 1
  fs, and in the beginning each wire is equilibrated for 0.1 ns.
Subsequently, the elongation process is simulated by separating the electrodes
at a constant velocity of 0.4~m/s. During this process the geometry is
recorded every 10~ps, in order to compute the transport properties. A
stretching process needs a total simulation time of about 4.5~ns until the
contact breaks.

\subsubsection{Tight-binding-based transport calculations}

The geometries obtained from the MD simulations are used to compute the
electronic contribution to the transport properties within the
Landauer-B\"uttiker formalism. In this case the electronic transmission that
determines both the electrical conductance and the electronic contribution to
the thermal conductance is computed with the help of a tight-binding model. To
be precise, we employ a non-orthogonal Slater-Koster tight-binding
parameterization, which has been constructed by fitting
  DFT-based results for the electronic band structure and total energies of
  metals across the periodic table, see
  Refs.~\onlinecite{Mehl1996,Papaconstantopoulos1998} for details. In this
parameterization we take into account the relevant valence orbitals, which for
Au and Pt include the $5d$, $6s$ and $6p$ orbitals, and for Al the $3s$, $3p$
and $3d$ orbitals. Moreover the hopping and overlap matrix elements in this
tight-binding model are functions of the distance between the atoms, which
enables us to use it with our MD simulations.

To compute the electronic transmission, we combine the tight-binding model
with NEGF techniques and the formulas detailed in section \ref{sec-LB}, very
much like in the DFT-based calculations. Details can be found in
Refs.~\onlinecite{Dreher2005,Pauly2006}. Briefly, as in the MD simulations,
the system is divided into three regions for the transport calculations, i.e.,
the two electrodes and the central wire, see Fig.~\ref{fig-geometry}. Because
the local environment of the atoms in the central part is very different from
that in the bulk, we impose a charge neutrality condition for all the atoms of
the central wire \cite{Dreher2005,Pauly2006}, which is known to be
approximately fulfilled in metallic systems. As in the DFT case the electrodes
are considered to be semi-infinite perfect crystals, and their surface Green's
functions are computed with the help of a decimation technique
\cite{Pauly2008,Guinea1983}.  Again, as in the DFT-based calculations, the
Green's function techniques also allow us to compute the individual
transmission coefficients $\tau_{\text{el},i}(E)$ at a given energy $E$.

\section{DFT-based transport results: Phonon transport} \label{sec-DFT-results}

\begin{figure}[t]
\begin{center} \includegraphics[width=\columnwidth,clip]{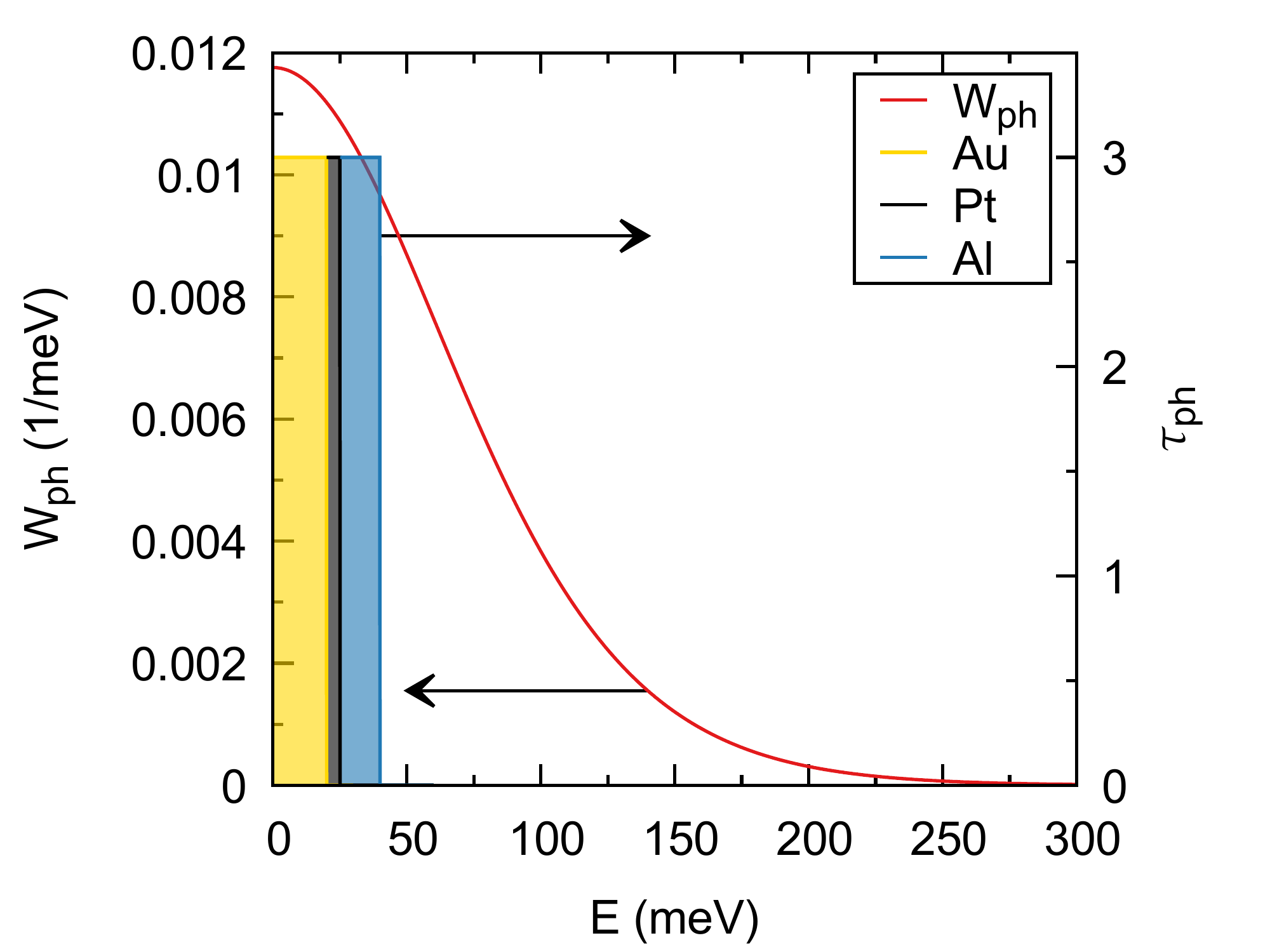} \end{center}
\caption{(Color online) Window function $W_{\rm ph}(E,T)$, defined in 
  Eq.~(\ref{eq-Wph}), as a function of energy at room temperature ($T=300$~K). 
  We also show the phonon transmissions used to estimate
  the phononic thermal conductance of single-atom contacts of Au, Pt and
  Al. We assume for simplicity that $\tau_{\text{ph}}(E)$ is equal to 3 for
  energies up to the Debye energy of the corresponding metal. Arrows in the 
  plot refer to left and right vertical scales.}
\label{fig-estimate}
\end{figure}

In this section we shall discuss the main results for the thermal conductance
of metallic atomic-size contacts, obtained with our ab initio, DFT-based
transport method. But before doing so, it is instructive to estimate the
contribution of phonons to the thermal conductance at room temperature. Let us
focus, in particular, on the case of single-atom contacts. To get estimates
for upper bounds, we assume that there are three phonon conduction
channels. We choose the number of three, since one expects one channel for
each spatial dimension. They might also be seen as a longitudinal and two
transverse eigenchannels. Let us furthermore assume that the channels exhibit
a perfect transparency for energies up to the corresponding Debye energy
$E_{\rm D}$ of the metal: 20 meV for Au, 25 meV for Pt, and 40 meV for
Al. Thus, if we use $\tau_{\rm ph}(E) = 3$ for $E \in [0, E_{\rm D}]$ in
Eq.~(\ref{eq-kph2}), as shown in Fig.~\ref{fig-estimate}, we obtain a room
temperature phononic thermal conductance of $0.199~\text{nW/K}=0.7 \kappa_0$
for Au, $0.244~\text{nW/K}=0.86 \kappa_0$ for Pt, and
$0.378~\text{nW/K}=1.33\kappa_0$ for Al. Notice that the largest value arises
for Al, which is simply due to its higher Debye energy as compared to Au and
Pt. These estimates need to be put into relation to the corresponding ones for
the electronic contribution to the thermal conductance. To obtain them, we use
the Wiedemann-Franz law, see Eq.~(\ref{eq-WF-law}), and the experimentally
reported values for the transmission of single-atom
contacts. We extract the transmissions from the lowest peak in
the electrical conductance histograms, which typically arises from
 single-atom contacts. In this way we expect a room temperature electronic
thermal conductance of about $0.568~\text{nW/K}=2\kappa_0$ for Au single-atom
contacts \cite{Agrait2003}, roughly between $0.710~\text{nW/K}=2.5\kappa_0$
and $1.420~\text{nW/K}=5\kappa_0$ for Pt
\cite{Nielsen2003,Smit2003,Evangeli2015}, and something between
$0.284~\text{nW/K}=\kappa_0$ and $0.568~\text{nW/K}=2\kappa_0$ for Al
\cite{Yanson1997,Cuevas1998b}. Thus, we see that the phononic and electronic
contributions could in principle be of similar order, and it is by no means
obvious that phonons can be ignored in the analysis of the heat conduction in
metallic atomic contacts. The quantitative determination of the relative
contributions of electrons and phonons to the thermal conductance will be a
central issue of the rest of this section.

\begin{figure*}[t]
\includegraphics[width=0.8\textwidth,clip]{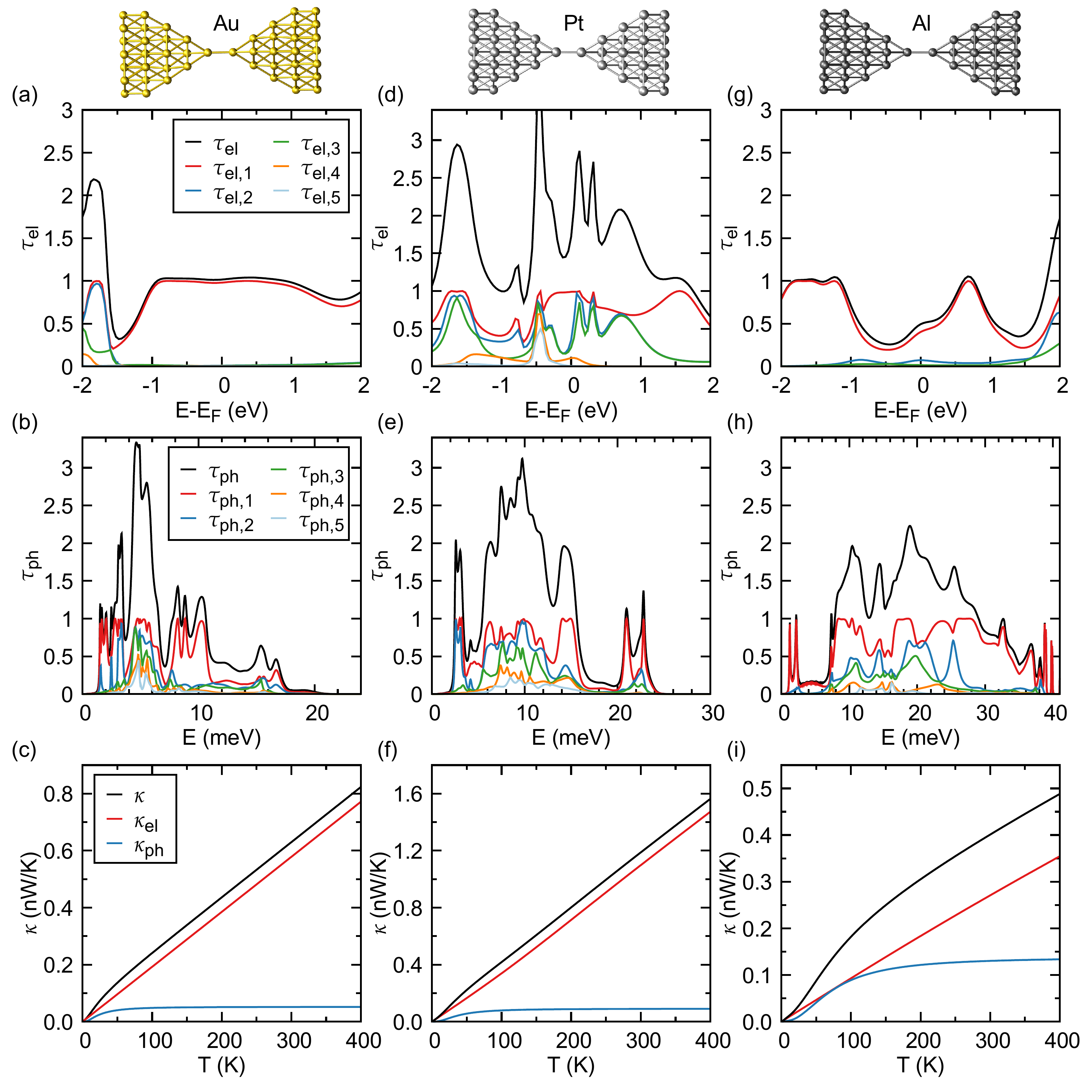}
\caption{(Color online) (a) Electronic transmission $\tau_{\rm el}$ as a
  function of energy (measured with respect to the Fermi energy $E_{\rm F}$)
  for the Au single-atom contact shown above the panel. We display the total
  transmission as well as the five largest transmission coefficients, as
  indicated in the legend. (b) The corresponding phononic transmission as a
  function of energy. Similar to the electronic transmission we show both the
  total one and the largest five individual transmission coefficients. (c)
  Thermal conductance as a function of temperature for the Au single-atom
  contact with the total thermal conductance $\kappa$ resolved into electronic
  and phononic contributions, $\kappa_{\rm el}$ and $\kappa_{\rm ph}$,
  respectively. (d-f) The same as in panels (a-c) for the Pt single-atom
  contact shown above panel (d). (g-i) The same as in panels
  (a-c) for the Al single-atom contact shown above panel (g).}
\label{fig-one-atom}
\end{figure*}
\begin{table*}[t]
\caption{\label{table1} Room temperature values of the different transport
  properties for the single-atom contacts of Fig.~\ref{fig-one-atom}: $G$ is
  the electrical conductance, $L_0TG$ is the expected result for the
  electronic contribution to the thermal conductance from the Wiedemann-Franz
  law, $\kappa_{\rm el}$ is the electronic thermal conductance, $\kappa_{\rm
    ph}$ is the phononic thermal conductance, and $\kappa$ is the total
  thermal conductance, determined as the sum of the electronic and phononic
  contributions.}
\begin{tabular}{ccccccc}
Material \; & \; $G$ ($G_0$) \; & \; $L_0TG$ (nW/K) \; & \; $\kappa_{\rm el}$ (nW/K) \; & 
\; $\kappa_{\rm ph}$ (nW/K) \; & \; $\kappa$ (nW/K) \; \\ \hline
Au & 1.01  & 0.577 & 0.578 & 0.051 & 0.629 \\
Pt & 1.83  & 1.041 & 1.098 & 0.098 & 1.196 \\
Al & 0.49  & 0.277 & 0.271 & 0.130 & 0.401 \\ \hline
\end{tabular}
\end{table*}

Let us start our discussion of the DFT-based transport results by considering
single-atom contacts of Au, Pt, and Al. Simulations, based on both classical
MD and DFT \cite{Jelinek2003,Dreher2005,Pauly2006}, show that the last
conductance plateau corresponds to one-atom-thick contacts that usually
feature an atomic dimer in the narrowest region before breaking. For this
reason we consider the ideal dimer geometries, shown in the upper part of
Fig.~\ref{fig-one-atom}, as representative examples of single-atom
contacts. For all the contacts the crystallographic $\langle 111 \rangle$
direction is oriented along the transport direction. Figure~\ref{fig-one-atom}
summarizes the transport results for the three junction materials under
study, and precise values of the relevant transport properties at room
temperature are summarized in Table~\ref{table1}. Fig.~\ref{fig-one-atom}(a)
displays the electronic transmission, both the total one and those of the five
highest transmission coefficients, as a function of energy for the Au dimer
contact. We find that the electronic transport around the Fermi energy is
dominated by a single conduction channel that is almost fully open, as it has
been reported before for similar geometries
\cite{Scheer1998,Cuevas1998a,Rubio-Bollinger2003,Pauly2008}. In this
particular case we find that the electrical conductance is $1.01G_0$, while
the corresponding result for the electronic thermal conductance at room
temperature is $\kappa_{\rm el} = 0.577~\text{nW/K}\approx 2\kappa_0$. This
value agrees very well with the expectation from the Wiedemann-Franz law in
Eq.~(\ref{eq-WF-law}), $L_0TG = 0.578~\text{nW/K}=2\kappa_0$. Turning now to
the phonon contribution to the thermal transport, we show in
Fig.~\ref{fig-one-atom}(b) the corresponding phononic transmission. In this
case the phononic transmission is mainly dominated by 3 conduction channels,
which are in general partially open. (Indeed the transmission values vary
between 0 and 1, depending on energy.) Notice that the transmission only
differs from zero below 20 meV, which corresponds to the Au Debye energy in
our calculations. Using this transmission function, we find that the phonon
contribution to the thermal conductance at room temperature is
$\kappa_{\rm ph} = 0.051~\text{nW/K} = 0.18\kappa_0$, which
is about 8\% of the total thermal conductance.
This value is much smaller than the upper bound provided above, because the
total transmission is clearly below 3 for almost all energies, which we
attribute to the mismatch between the incoming phonons and the local
vibrations in the narrowest part of the contact. For completeness we also
present in Fig.~\ref{fig-one-atom}(c) the temperature dependence of the
thermal conductance, including the electronic and phononic contributions, as
well as the total one. Note that hereafter $\kappa=\kappa_{\rm el} + 
\kappa_{\rm ph}$ denotes the total thermal conductance. As one can see,
and we already explained in Ref.~\onlinecite{Cui2017}, the thermal conductance
of Au dimer contacts is clearly dominated by electrons for most
temperatures. On the other hand, the room temperature thermal conductance
quantization, as observed in Ref.~\onlinecite{Cui2017}, is a consequence of
the fact that, in addition, the electrical conductance is quantized and the
electronic transmission is rather smooth around the Fermi energy, see
Fig.~\ref{fig-one-atom}(a), which implies that the Wiedemann-Franz law is
accurately fulfilled. Note that we will refer to electrical conductance
quantization as a situation, where all open transmission channels exhibit
perfect transparency, i.e. $\tau_{\text{el},i}(E_{\text{F}})$ is either 1 or
0. Let us furthermore recall that the tendency of Au single-atom contacts to
exhibit an electrical conductance of around $1G_0$ is due to the fact that the
electronic transport is dominated by the $s$ valence orbitals of this
metal. In general the number of conduction channels in a single-atom contact
is determined by the number of valence orbitals that give a significant
contribution to the density of states around the Fermi energy
\cite{Cuevas1998a,Scheer1998}.

We will now discuss the results for the Pt single-atom contact shown at the
top of the second column in Fig.~\ref{fig-one-atom}. The
corresponding total electronic transmission function is displayed in
Fig.~\ref{fig-one-atom}(d) along with the five most relevant transmission
coefficients. As one can see, and in strong contrast with the Au case,
there are four conduction channels that provide a sizable contribution to the
transport at the Fermi energy. This is due to the fact that apart from the $s$
valence orbitals, the $d$ valence orbitals of Pt atoms also contribute to the
electronic transport \cite{Sirvent1996,Pauly2006,Evangeli2015}. Moreover these
orbitals yield conduction channels that are partially open, which naturally
explains the lack of electrical conductance quantization in this metal
\cite{Nielsen2003,Smit2003,Evangeli2015}. The transmission function results in
an electrical conductance of $1.83G_0$, while the corresponding electronic
contribution to the thermal conductance at room temperature is
$1.098~\text{nW/K}=3.87\kappa_0$. This is close to the value of
$1.041~\text{nW/K}=3.67\kappa_0$ suggested by the Wiedemann-Franz law. The
larger deviation from this law, as compared to Au, arises from the more
pronounced energy dependence of the electronic transmission function of Pt
around the Fermi energy, caused by the $d$ bands
\cite{Pauly2006,Pauly2011,Evangeli2015}. In contrast to the electronic
transport, the shape of the phonon transmission of this Pt contact is similar
to that of the Au contact, as visible from Fig.~\ref{fig-one-atom}(e). This
originates from the similar masses of Au and Pt atoms, leading to comparable
Debye energies. As in the case of Au, three to four conduction channels
dominate the phonon transport in the Pt dimer contact, leading to a phonon
thermal conductance of $0.098~\text{nW/K}=0.35\kappa_0$. This value almost
doubles $\kappa_{\text{ph}}$ of the Au contact, but is comparable in relative
terms: It also constitutes about 8\% of the total thermal conductance $\kappa$
(see Table~\ref{table1}). Fig.~\ref{fig-one-atom}(f) shows that, very much
like in the case of Au, the thermal conductance is largely dominated by the
electrons at all relevant temperatures. The lack of electrical conductance
quantization results in the absence of thermal conductance quantization for
single-atom contacts of Pt, as was confirmed experimentally in
Ref.~\onlinecite{Cui2017}.

Let us now address the Al single-atom contact displayed above
Fig.~\ref{fig-one-atom}(g). We remark that the thermal transport in Al
contacts has not been investigated experimentally so far. Aluminum is a
reactive metal that is not easy to handle at room temperature, but the
electronic transport through Al atomic contacts has been thoroughly explored
at low temperatures. It is a very good example of a light metal with a Debye
energy that is significantly larger than those of Au and Pt. The total
electronic transmission around the Fermi energy stems from two to three
partially open channels, as one can see in Fig.~\ref{fig-one-atom}(g) and has
been reported both theoretically and experimentally in numerous occasions
\cite{Scheer1997,Scheer1998,Cuevas1998a,Cuevas1998b,Jelinek2003,Schirm2013}. These
observations can be understood by the contribution of both $s$ and $p$ valence
orbitals of Al atoms \cite{Scheer1998,Cuevas1998a}. The electrical conductance
in this example is $0.49G_0$, while the room temperature electronic thermal
conductance is $0.277~\text{nW/K}=0.98\kappa_0$, in very good agreement with
the value of $0.271~\text{nW/K}=0.95\kappa_0$ from the Wiedemann-Franz law. As
one can see at the phonon transmission function, displayed in
Fig.~\ref{fig-one-atom}(h), the phonon transport is also dominated by three to
four conduction channels, like in the Au and Pt cases, but now
phonon modes up to 40 meV participate. These additional phonon modes give
rise to a phonon thermal conductance of $0.130~\text{nW/K}=0.46\kappa_0$. This
value is clearly larger than those of Au and Pt. Together with the lower
electrical conductance, $\kappa_{\text{ph}}$ yields now about to 32\% of
$\kappa$ (see also Table~\ref{table1}). Figure~\ref{fig-one-atom}(i) shows the
corresponding temperature dependence of the thermal conductance for the Al
dimer contact. Notice that contrary to the cases of Au and Pt the phonon
contribution is now of the same size as the electronic one in a broad range of
temperatures up to $T=100$~K and at room temperature it still constitutes a
very significant contribution. These results suggest that a clear violation of
the Wiedemann-Franz law should be observable in Al single-atom contacts, a
prediction that yet awaits experimental verification.

\begin{figure}[t]
\includegraphics[width=\columnwidth,clip]{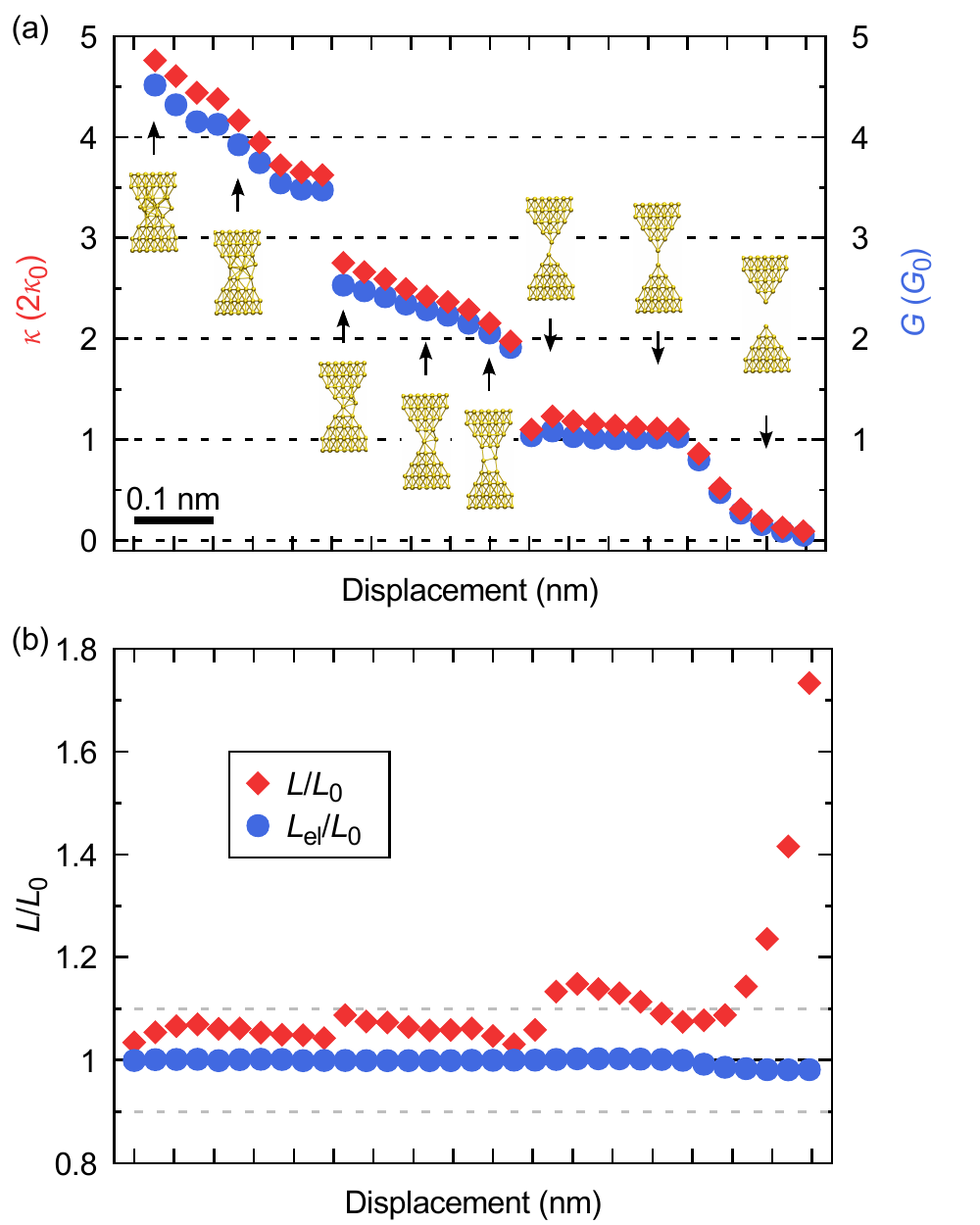}
\caption{(Color online) (a) Electrical and thermal conductance at room
  temperature as a function of electrode displacement for an Au contact,
  oriented along the fcc $\langle 111 \rangle$ crystallographic direction. The
  blue dots correspond to the results for the electrical conductance (right
  vertical scale), which is normalized by the electrical conductance quantum
  $G_0$. The red diamonds correspond to the total thermal conductance (left
  vertical scale), taking into account both the electronic and phononic
  contributions, and it is normalized by twice the thermal conductance quantum
  $\kappa_0$. The different geometries, shown in this panel, correspond to
  snapshots taken during the elongation and compression processes. (b)
  The corresponding Lorentz ratios, defined via Eq.~(\ref{eq-Lorentz-ratio}). 
  The red diamonds show the full ratio computed with the total thermal 
  conductance, while the blue dots show the electronic Lorentz ratio, if the 
  thermal conductance consists only of the electronic contribution.}
\label{fig-Au-111}
\end{figure}
\begin{figure}[t]
\includegraphics[width=\columnwidth,clip]{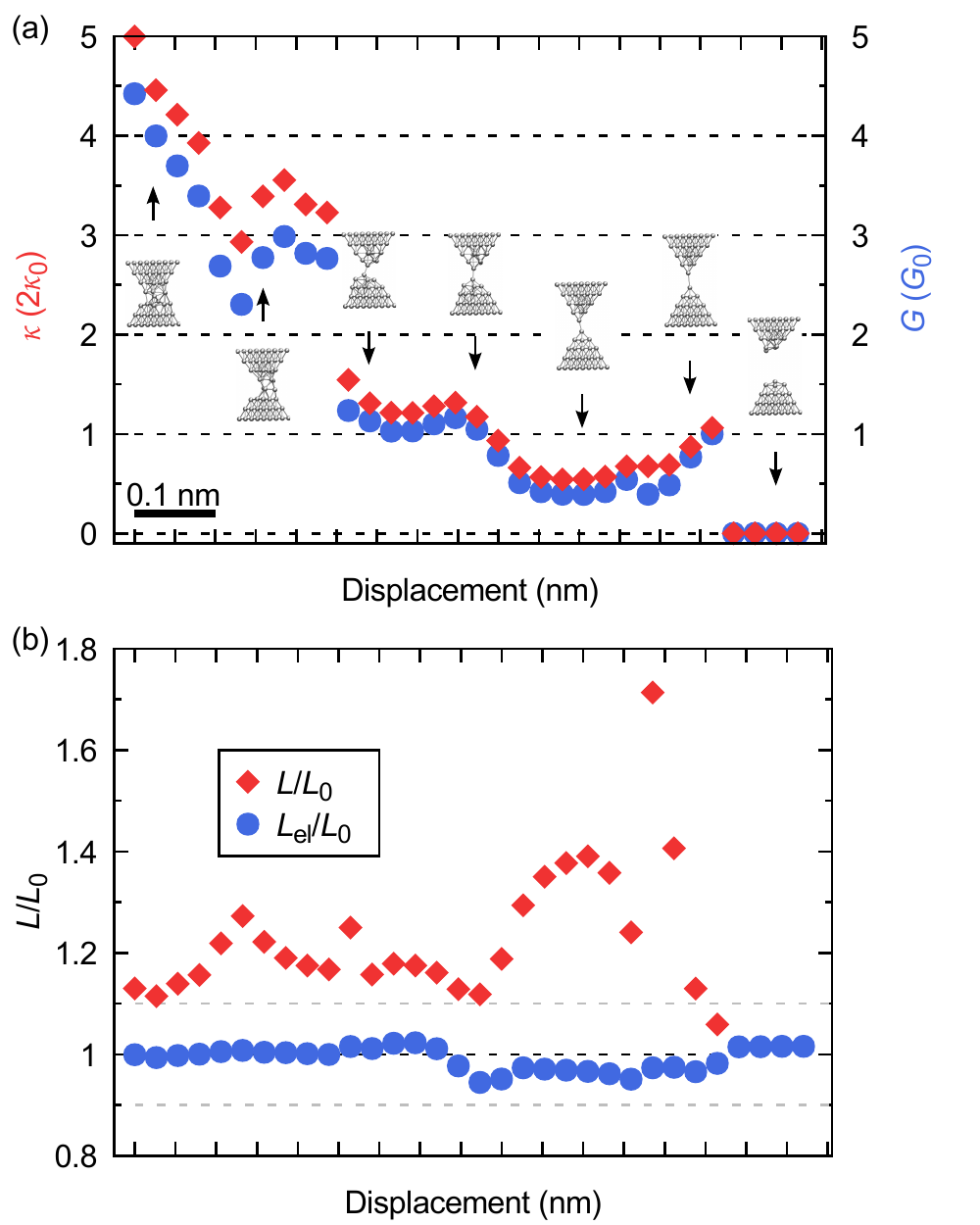}
\caption{(Color online) The same as in Fig.~\ref{fig-Au-111}, but for an Al
  contact.}
\label{fig-Al-111}
\end{figure}

To further study the relative contributions of electrons and phonons to the
thermal conductance and to model the experiments more closely, we have used
our DFT-based approach to simulate the contact formation and to compute the
corresponding conductance traces. Since Au and Pt behave similarly with regard
to the small contribution of phonons to $\kappa$, we concentrate in the
following on a comparison of Au and Al only. We start out with
atomic contacts of Au and Al, and stretch or compress the geometry
adiabatically. This is done by separating or by approaching the electrodes
in a step-like manner and by subsequent re-optimization of the junction
geometry. As displacement step we use 0.26~\AA\ and compute $G$,
$\kappa_{\text{el}}$, $\kappa_{\text{ph}}$ and $\kappa$ with our DFT-based
transport method for the obtained equilibrium geometries.

In Fig.~\ref{fig-Au-111}(a) we show the results of such a
simulation for the Au contact of Fig.~\ref{fig-one-atom} that is grown along
the $\langle 111 \rangle$-direction of the fcc lattice. We started with the
dimer contact represented by the second geometry in that panel, counting
from the right. This geometry was compressed to obtain thicker cross
sections as well as stretched to simulate the breaking of the contact. In
Fig.~\ref{fig-Au-111}(a) we show the results for $G$ and $\kappa$ at room
temperature for the series of contacts, obtained following our protocol. The
electrical conductance in this plot is normalized by the electrical
conductance quantum, $G_0$, while the thermal conductance is normalized by
$2\kappa_0$ with $T=300$~K. As one can see, both conductances proceed in a
step-like manner in a succession of plateaus and abrupt jumps, related to
elastic stages, where bonds are stretched and forces build up, and plastic
stages, where bonds break and the accumulated tension is released. This
behavior resembles the experiments \cite{Cui2017}. In addition both
conductances follow each other very closely. With the normalization used here,
this means that the Wiedemann-Franz law is well obeyed. Notice also that both
$G$ and $\kappa$ feature a plateau at $1G_0$ and $2\kappa_0$, respectively,
which illustrates the tendency of Au single-atom contacts to exhibit quantized
electronic and thermal transport, even at room temperature \cite{Cui2017}.

To quantitatively assess the validity of the Wiedemann-Franz law, it is
customary to define the so-called Lorentz ratio $L/L_0$ as follows
\begin{equation}
  \frac{L}{L_0} = \frac{\kappa}{L_0TG} .
  \label{eq-Lorentz-ratio}
\end{equation}
Here, $\kappa=\kappa_{\text{el}}+\kappa_{\text{ph}}$ is the total thermal
conductance due to both electrons and phonons, $L_0$ is the Lorentz number,
and $G$ is the electrical conductance. A Lorentz ratio equal to 1 means that
the measured thermal conductance agrees exactly with the expectations from the
Wiedemann-Franz law, while deviations from 1 signal that this relation is
violated. Such violations could be due to the contribution of phonons or they
can have an electronic origin. In Fig.~\ref{fig-Au-111}(b) we show the Lorentz
ratio (red diamonds), using the results of panel (a). As one
can see, there are small deviations from 1 on the order of 5-10\% in the
contact regime, depending on the exact junction geometry. Theses observations
are in good agreement with the measurement of Ref.~\onlinecite{Cui2017}. In
order to better understand the origin of these deviations, we also show in
Fig.~\ref{fig-Au-111}(b) (blue dots) the electronic Lorentz ratio
$L_{\text{el}}/L_0=\kappa_{\text{el}}/(L_0TG)$, constructed by replacing
$\kappa$ with $\kappa_{\text{el}}$ in Eq.~(\ref{eq-Lorentz-ratio}). As one can
see, this electronic Lorentz ratio is very close to 1, irrespectively of the
electrode displacement. This means that the deviations from the
Wiedemann-Franz law are mainly due to phonons. It is also worthwhile to
consider in detail the behavior of $L/L_0$ with the electrode displacement in
Fig.~\ref{fig-Au-111}(b). Deviations from the Wiedemann-Franz law tend to
increase with larger displacements, i.e.\ towards smaller minimal contact
cross sections. Furthermore $L/L_0$ exhibits a sawtooth-like shape, typically
decreasing within each elastic stage. Taking into account that
$L_{\text{el}}/L_0\approx 1$ implies $L/L_0\approx
\kappa/\kappa_{\text{el}}=1+\kappa_{\text{ph}}/\kappa_{\text{el}}$, this
means that the relative weight of $\kappa_{\text{ph}}$ in $\kappa$ tends to
reduce with increasing stress in the Au single-atom contacts. We attribute
this to overall decreasing force constants with pulling or, in other words, a
softening of interatomic bonds. Revivals of $L/L_0$ are seen at the points,
where bonds break and the atomic contact reconfigures.

In Fig.~\ref{fig-Al-111} we show the results of the
simulation for an Al atomic wire that was performed following exactly the
same protocol as in the Au simulation of Fig.~\ref{fig-Au-111}, where we use
the Al dimer contact of Fig.~\ref{fig-one-atom} as starting geometry. As in
the Au case, the electrical and thermal conductances proceed in a step-like
manner with the peculiarity that most plateaus exhibit a positive slope at the
end of each plateau, i.e., the conductance increases upon stretching before
bonds break. This unique behavior of Al contacts is well-known, and it has
been observed in different experiments and convincingly explained
\cite{Scheer1997,Scheer1998,Cuevas1998b,Jelinek2003}. The electrical and
thermal conductance are correlated, but larger deviations as compared to Au
are visible.  This is well apparent in the Lorentz ratio, shown in
Fig.~\ref{fig-Al-111}(b), which features deviations from the Wiedemann-Franz
law as large as 40\% and even above. Notice also that the electronic
contribution to the thermal conductance follows closely the prediction of the
Wiedemann-Franz law, with deviations that are at most about 5\%, as can be
inferred from the electronic Lorentz ratio $L_{\text{el}}/L_0$. Thus, the
larger violations of the Wiedemann-Franz relation that we find for Al are
mainly due to the phonon contribution to the thermal transport.  Our results
illustrate that the phonon thermal conductance cannot always be neglected,
when analyzing the thermal transport of metallic atomic-size contacts. The
sawtooth-like behavior of $L/L_0$, discussed for Au before, is not so apparent
for this Al contact. In any case the Lorentz ratio exhibits minima close to
the displacement values, at which atomic bonds break.

\begin{figure}[t]
\includegraphics[width=\columnwidth,clip]{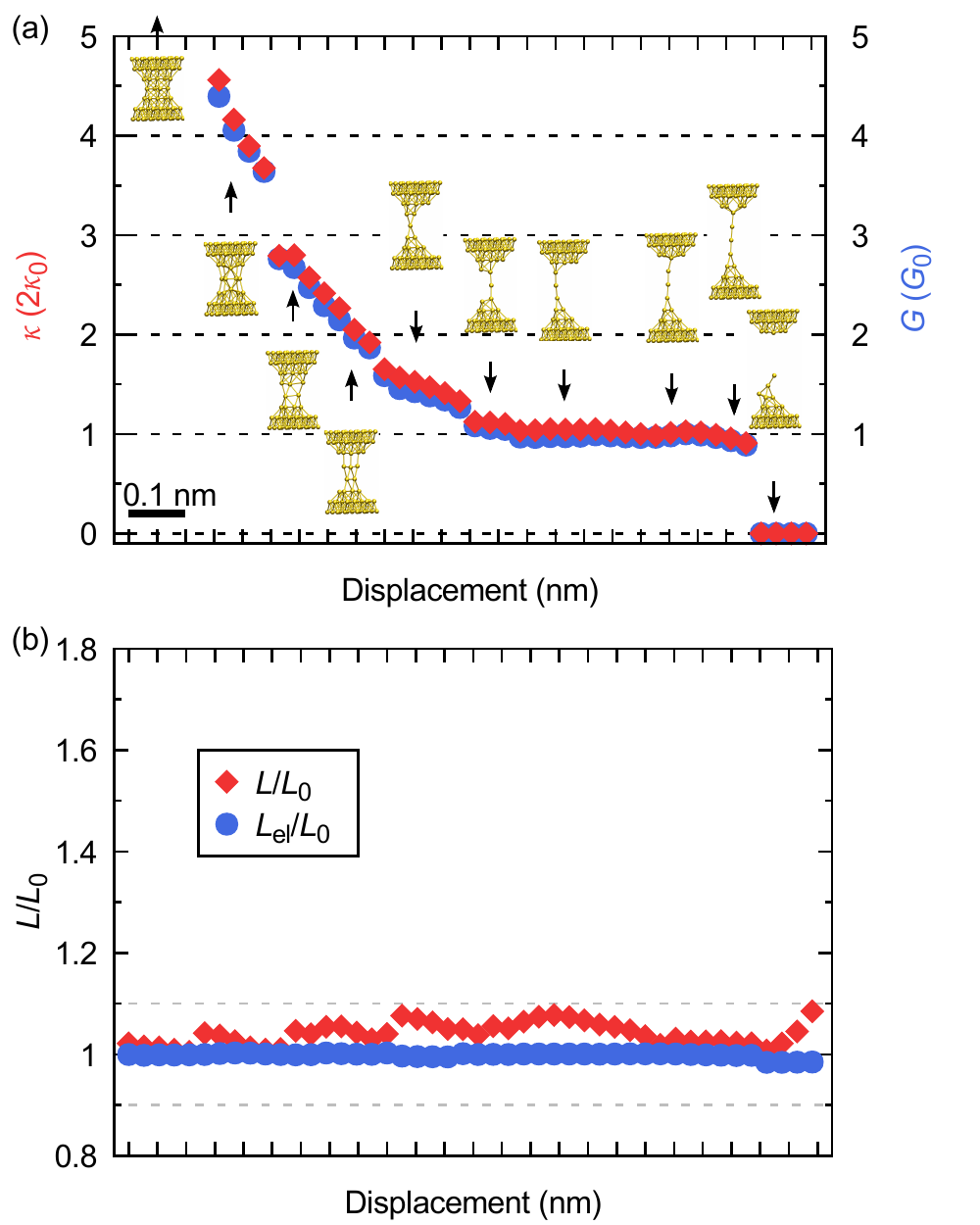}
\caption{(Color online) The same as in Fig.~\ref{fig-Au-111}, but for an Au
  contact grown along the fcc $\langle 100 \rangle$ crystallographic
  direction. The contact was elongated, starting with the geometry shown in
  the upper left part of panel (a).}
\label{fig-Au-100}
\end{figure}
\begin{figure}[t]
\includegraphics[width=\columnwidth,clip]{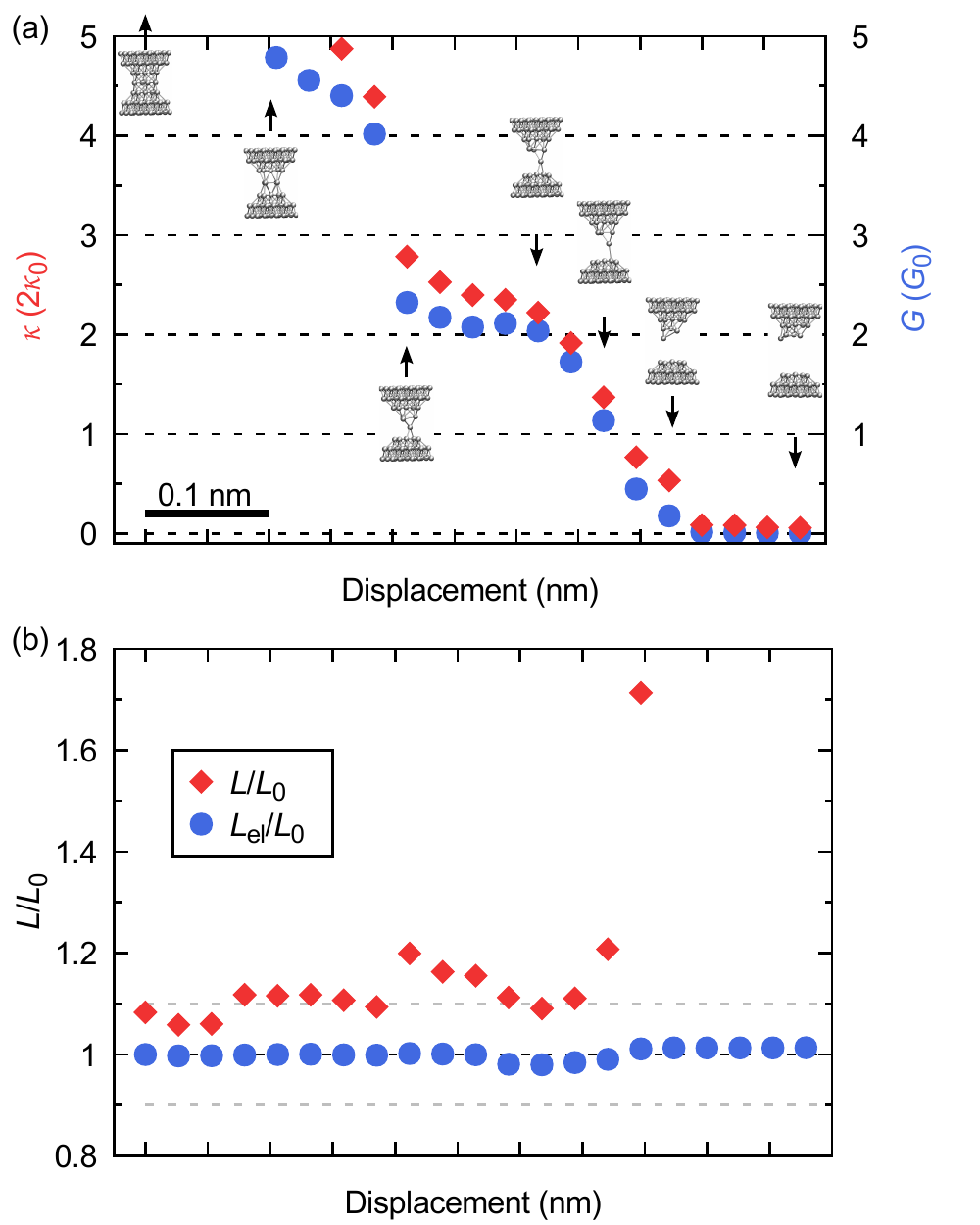}
\caption{(Color online) The same as in Fig.~\ref{fig-Au-100}, but for an Al
  contact.} \label{fig-Al-100}
\end{figure}

It is worth stressing that we have checked that the conclusions above are not
an artifact of the protocol used to simulate the contact formation or the
choice of the crystallographic direction of the contact geometries. In
Fig.~\ref{fig-Au-100} we show an example of such tests, where we have
simulated the contact formation of another Au atomic wire. In this case the
contact is grown along the fcc $\langle 100 \rangle$-direction, and we started
the simulation with the geometry shown in the upper left part of
Fig.~\ref{fig-Au-100}(a). As before, we then stretched the contact
progressively in steps of 0.26~\AA. We find that all the
basic observations made above about the Au contacts are reproduced
here. Notably, this Au wire forms a chain with up to four atoms in length in
the last stages before breaking. During the formation of this atomic chain,
$G$ and $\kappa$ remain approximately quantized with values of $1G_0$ and
$2\kappa_0$, respectively. The formation of such Au atomic chains has been
reported in numerous experiments
\cite{Smit2003,Yanson1998,Ohnishi1998,Rodrigues2000}, and their electronic
transport properties have been amply discussed in the literature
\cite{Agrait2003,Cuevas2017}. As for the Au contact before, the plots of the
Lorentz ratio show that the relative contribution of phonons to the total
thermal conductance tends to decrease with increasing tension in the contact.

As a last example, we briefly discuss the stretching simulation for the Al
contact depicted in Fig.~\ref{fig-Al-100}. As in the Au case of
Fig.~\ref{fig-Au-100}, this Al contact is grown along the fcc $\langle 100
\rangle$-direction and the stretching simulations were initiated with the
geometry displayed in the upper left part of Fig.~\ref{fig-Al-100}(a). There
are two features of the results in Fig.~\ref{fig-Al-100} that we want to
highlight. First of all, the last plateau before the breaking of the wire
exhibits an electrical conductance of about $2G_0$. It originates from a
one-thick contact, where there is a monomer in the narrowest region,
as opposed to the dimer realized before breaking in the
previous Al simulation of Fig.~\ref{fig-Al-111}. As we shall argue in the
next section, geometries of this type are responsible for a second peak in the
electrical conductance histogram of this metal close to $2G_0$. The second
feature worth remarking is that, as one can see in Fig.~\ref{fig-Al-100}(b),
the phonons give a slightly smaller contribution to the total thermal
conductance, as compared with the example of Fig.~\ref{fig-Al-111}, but still
clearly larger than in the Au case. The sawtooth-like shape of $L/L_0$ as a
function of electrode displacement in Fig.~\ref{fig-Al-100}(b) is much more
pronounced than in Fig.~\ref{fig-Al-111}.

Throughout this section, we have so far only discussed the contact regime of
heat transport, which has been studied experimentally
\cite{Mosso2017,Cui2017}.  But it is also interesting to examine the tunneling
regime, when contacts are broken. This regime is still
challenging for experimentalists, since signals are rather small and
contamination of surfaces may play a crucial role \cite{Cui2017a}.
Moreover, as the gap between the electrodes increases, at some point the
thermal radiation via photon tunneling may give a contribution of similar
order \cite{Klockner2017a}.  Figures~\ref{fig-Au-111}, \ref{fig-Al-111},
\ref{fig-Au-100} and \ref{fig-Al-100} consistently show that in the tunneling
regime the electronic Lorentz ratio $L_{\text{el}}/L_0$ stays close to
1. However, $L/L_0$ typically increases significantly in our calculations,
meaning that the phonon thermal conductance decays more slowly with distance
than the electronic part. This can be rationalized by considering that the
electronic thermal conductance arises only from the overlap of electronic
wavefunctions, which decreases exponentially with the separation of left and
right junction parts. Lattice vibrations may instead couple over longer
distances, because multipolar electrostatic charge contributions can for
instance lead to an algebraic decay of $\kappa_{\text{ph}}$
\cite{Xiong2014,Chiloyan2015}.

\section{Electronic thermal conductance: Statistical test of the Wiedemann-Franz law}\label{sec-TB-results}

\begin{figure*}[t]
\includegraphics[width=0.9\textwidth,clip]{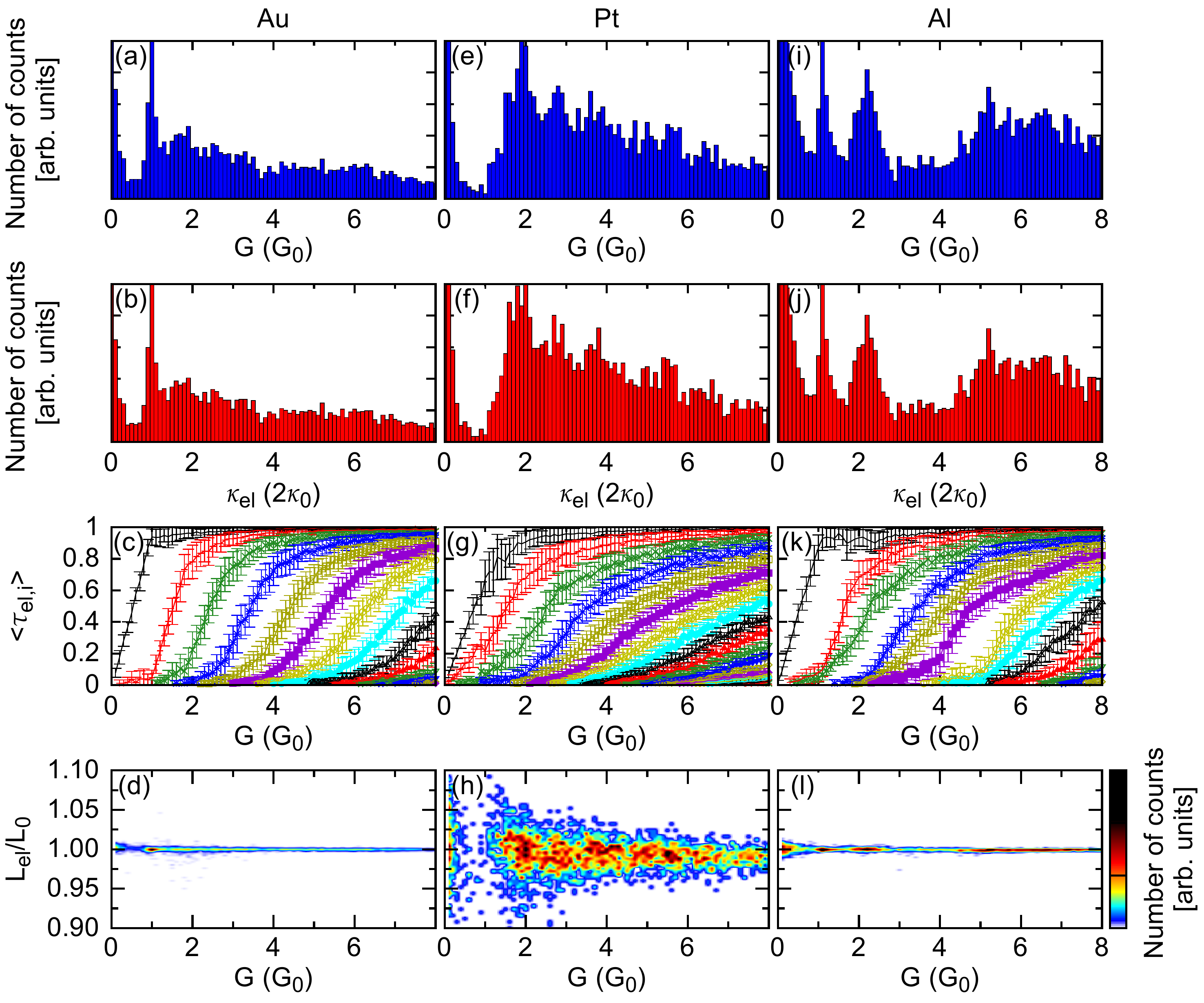}
\caption{(Color online) (a) Electrical conductance histogram obtained from 100
  MD simulations of the stretching of Au atomic contacts at room
  temperature. (b) The corresponding electronic thermal conductance
  histogram. (c) The 20 largest electronic transmission coefficients as a
  function of conductance for the Au simulations. The lines correspond to the
  average values and the bars to the standard deviations. (d) Density plot of
  the electronic Lorentz ratio as a function of electrical conductance for the
  Au contacts. (e-h) The same as in panels (a-d), but for Pt contacts. (i-l)
  The same as in panels (a-d), but for Al contacts.}
\label{fig-histo}
\end{figure*}

The DFT simulations presented in the previous section represent the state of
the art in the modeling of the thermal transport of atomic-scale
systems. Ideally, one would like to perform many such simulations for even
larger contacts to describe the conductance histograms for $G$ and $\kappa$
that are reported experimentally. However, these simulations are very
time-consuming, and a study of conductance histograms with DFT-based methods
is presently not feasible for us. Instead we employ an alternative method
based on the combination of classical MD simulations and quantum transport
calculations based on a tight-binding model. As described in section
\ref{sec-MD}, this hybrid approach allows us to determine the geometries,
realized in the experiments, as well as to compute the electronic transport,
i.e.\ $G$ and $\kappa_{\text{el}}$. In particular, as shown by us previously
\cite{Dreher2005,Pauly2006,Evangeli2015,Cui2017}, the approach allows us to
calculate conductance histograms and, in turn, to establish a very direct
comparison with experiments. As compared to the adiabatic
stretching processes at $T=0$, assumed in our DFT simulations above, the MD
simulations offer the advantage to take also the influence of finite
temperature on the formation of junction geometries into account. In this
work we have used the hybrid MD and tight-binding approach to study in a
systematic manner the electronic contribution to the thermal conductance in
atomic contacts in order to test, whether deviations from the Wiedemann-Franz
law are expected from purely electronic effects.

In Fig.~\ref{fig-histo} we summarize the results, obtained from 100 MD
simulations of the formation of Au, Pt and Al atomic contacts at room
temperature. In this figure we show the electrical conductance histograms, the
electronic thermal conductance histograms, the evolution of electronic
transmission coefficients with conductance, and the electronic Lorentz
ratio. Let us stress that we have checked that 100 simulations are enough to
converge the main features of the conductance histograms. We also want to
point out that the results for Au and Pt were already presented in
Ref.~\onlinecite{Cui2017} and are shown here for comparison with Al, which has
not been analyzed before. The first thing to notice in Fig.~\ref{fig-histo} is
the good correlation between the histograms of the electrical conductance and
those of the electronic thermal conductance, which clearly indicates that the
Wiedemann-Franz law is nicely fulfilled, where we ignore here of course the
contribution of phonons. This is more apparent in the lower panels, where we
show the electronic Lorentz ratio. As one can see, for the cases of Au and Al
the deviations from 1 are very small. They are on the order of 1-2\%, while in
the case of Pt they are slightly larger and can reach up to around 5\%, which
is consistent with our DFT results. As discussed above, these larger deviations 
in the case of the transition metal Pt are due to a more pronounced energy dependence 
of the electronic transmission around the Fermi energy, which in turn is due to 
the fact that atomic $d$ valence orbitals play a major role in the electronic
transport of this metal. In summary our results show that no significant
deviations from the Wiedemann-Franz law are expected from purely electronic
effects, irrespective of the material or the contact size.

With respect to the conductance histograms there are several features that we
want to emphasize.  First of all, the Au histograms are dominated by a large
peak around $1G_0$ for the electrical conductance and around $2\kappa_0$ for
the electronic thermal conductance. Since in the Au case the phonons play a
relatively minor role, this explains the observed quantized
transport of $\kappa$ in Ref.~\onlinecite{Cui2017}. The large peak for $G$
and $\kappa_{\text{el}}$ in the Au histograms originates from single-atom
contacts, where the electronic transport is dominated by a single, fully open
channel, as can be deduced from Fig.~\ref{fig-histo}(c). The corresponding
histograms for Al are very interesting. They exhibit two peaks close to
quantized values at $1G_0$ and $2G_0$, which may give the impression that the
electronic transport is also quantized in Al few-atom contacts. However, this
is actually not the case in the sense of our definition in
section~\ref{sec-DFT-results}, which requires that transmission coefficients
$\tau_{\text{el},i}$ are either 1 or 0. As we show in Fig.~\ref{fig-histo}(g),
the main contribution to those two peaks comes from contacts, where at least
three partially open channels give a significant contribution to the
transport. A detailed analysis shows that the lowest peak close to $1G_0$ for
the electrical conductance originates from one-atom-thick contacts, featuring
a dimer in the narrowest part, while the second peak around $2G_0$ stems
mainly from contacts with only a single atom at the tightest
constriction. This is indeed consistent with the observations of our DFT
simulations in Figs.~\ref{fig-Al-111} and \ref{fig-Al-100}. The fact that the
transmission coefficients of the open conduction channels in the last two
plateaus of Al wires add up to a total transmission close to integer values
is, in general, merely a coincidence. We do find, however, that there is a
tendency for Al contacts to break, when the electrical conductance is close to
$1G_0$, like in the DFT simulation of Fig.~\ref{fig-Al-111}. In those cases
the transport is indeed dominated by a single conduction channel. However,
along the last plateaus this is normally not the case, but we find that at
least two additional channels contribute to the electronic transport. Let us
remark that the multi-peak structure in the electrical conductance histogram
of Al contacts has been reported experimentally \cite{Yanson1997}, and it is
qualitatively reproduced by our MD simulations. Let us also stress that our
interpretation of the lack of quantized electronic transport in Al contacts
has been verified in great detail in experiments, where the electronic
conduction channels were determined with the help of superconductivity
\cite{Scheer1997,Scheer1998,Schirm2013}. Finally, in the Pt case the histogram
only features a rather broad peak slightly below $2G_0$ for the electrical
conductance. This broad peak and the lack of quantized transport are due to
the fact that even in the case of Pt single-atom contacts several conduction
channels with intermediate transmissions give a significant contribution to
the electronic transport, see Fig.~\ref{fig-histo}(k). As explained in the
previous section, this is due to the electronic structure of the transition
metal Pt, in which the atomic $d$ valence orbitals contribute
decisively to both the density of states around the Fermi energy and to the
electronic transport.

\section{Conclusions} \label{sec-Conclusions}

In summary, motivated by recent experiments \cite{Cui2017,Mosso2017}, we have
used state-of-the-art theoretical techniques to perform a systematic study of
the thermal conductance of metallic atomic-size contacts made of Au, Pt and
Al. In particular we have investigated two points of special
interest, namely the contribution of phonons to the thermal transport and the
validity of the Wiedemann-Franz law.  Our first-principles transport
calculations based on DFT show that in the case of heavy metals like Au and
Pt, which were experimentally investigated in Ref.~\onlinecite{Cui2017},
phonons provide a modest contribution to the thermal conductance (typically
below 10\%), and this conductance is given, to a very good approximation, by
the Wiedemann-Franz law. Moreover our calculations show that, while the
thermal conductance of Au single-atom contacts is quantized due to the fact
that the electronic transport is dominated by a single, fully open channel, in
the case of Pt such a quantization is not present, since additional electronic
conduction channels originating from the Pt $d$ bands contribute. In the case
of Al, a light metal with a much higher Debye energy, we find that the
relative contribution of phonons to the thermal transport is considerably
larger than for Au and Pt. Indeed, it can be as large as 40\% of the total
thermal conductance, depending on the contact geometry. This is primarily 
caused by the fact that, because of the higher Debye energy of Al, the phonon 
modes yield a larger contribution to the heat transport, but also due to the 
fact that the electrical conductance of Al junctions is somewhat lower than 
those of Au or Pt for the same contact size. Thus, our calculations show that, 
in general, the phonon transport cannot be ignored, when evaluating the thermal
conductance of metallic atomic-size contacts. This is clearly at variance with
the case of macroscopic wires, in which the phonon contribution to thermal
transport is basically negligible, irrespective of the chosen (standard)
metal.

Beyond the contact regime, our DFT simulations predict
that phonons will contribute substantially more to thermal transport than
electrons, when junctions are in the tunneling regime. While this aspect was
not studied systematically, a possible explanation might be that
electrostatic interactions between dipoles or higher multipoles of the
electrode geometry dominate for large gaps, where the overlap of electronic
wavefunctions of the broken sides of the contacts has decayed exponentially.

On the other hand, since our DFT results and the experiments \cite{Cui2017}
consistently suggest that electrons are more important in heat transport than
phonons for metallic atomic-size contacts in the contact regime, we have
employed classical MD simulations and a tight-binding model to carry out a
statistical study of the purely electronic transport. In particular, we have
analyzed, whether the electronic contribution to the thermal conductance
follows the Wiedemann-Franz law. Our results for Au, Al and Pt atomic wires
show that only minor deviations from the Wiedemann-Franz law are expected from
the electronic contribution, irrespective of the contact size. The largest
deviations, albeit modest (below 5\%), are obtained in transition metals like
Pt, where the energy dependence of the electronic transmission is more
pronounced than for Au and Al due to the contribution of the $d$
orbitals. This energy dependence is also reflected in other
  transport properties such as the thermopower, which has been found to be
  larger in single-atom contacts of Pt than of Au \cite{Evangeli2015}. On the
  other hand, the fact that the Wiedemann-Franz law is accurately fulfilled in
  these metallic nanowires provides a practical way to estimate the electronic
  thermal conductance from the knowledge of the electrical conductance. This,
  in turn, enables the extraction of the phonon contribution from the
  experimental results of the thermal conductance.

Overall our results provide deep insight into the thermal transport in
metallic atomic-size contacts, one of the most important testbeds for
nanoscale transport. Our results, in turn, may prove important for related
systems such as single-molecule junctions, whose thermal transport properties
should soon be amenable to measurement with the very same techniques that have
finally enabled the exploration of the thermal conductance of metallic atomic
contacts \cite{Cui2017,Mosso2017}.

\section{Acknowledgments}

We are indebted to Pramod Reddy, Edgar Meyhofer and Longji Cui for numerous
discussions on the experimental results of Ref.~\onlinecite{Cui2017}.
J.C.K.\ and F.P.\ were financially supported by the Carl Zeiss Foundation, the
Collaborative Research Center (SFB) 767 of the German Research Foundation
(DFG), and the Junior Professorship Program of the Ministry of Science,
Research, and the Arts Baden-W\"urttemberg.  M.M.\ and P.N.\ acknowledge
funding from the SFB 767 and computer time granted by the John von Neumann
Institute for Computing. J.C.C.\ thanks the Spanish Ministry of Economy and
Competitiveness (Contract No.\ FIS2014-53488-P) for funding as well as the DFG
and SFB 767 for sponsoring his stay at the University of Konstanz as Mercator
Fellow. An important part of the numerical modeling was carried out on the
computational resources of the bwHPC program, namely the bwUniCluster and the
JUSTUS HPC facility.


\end{document}